\documentclass[aps,prd,a4paper,reprint,superscriptaddress,floatfix,nofootinbib,amsmath,amssymb]{revtex4-2}
\pdfoutput=1   
\usepackage{amsmath,amsfonts,amssymb,amsbsy}
\usepackage{mathrsfs,latexsym}
\usepackage{graphicx}
\usepackage{xcolor}
\usepackage[bookmarks,
bookmarksopen = true,
bookmarksnumbered = true,
linktocpage,
colorlinks = true,
linkcolor = blue,
urlcolor  = blue,
citecolor = blue,
anchorcolor = green,
hyperindex = true,
hyperfigures]{hyperref}
\usepackage{array}
\usepackage{booktabs}
\usepackage{multirow}
\usepackage{subfig}
\usepackage{placeins}
\usepackage{blindtext}
\usepackage{appendix}
\usepackage{bm}
\usepackage{times}
\usepackage{float}
\usepackage{booktabs}
\usepackage{wrapfig}
\usepackage[utf8]{inputenc} 
\usepackage[T1]{fontenc} 
\usepackage{amsmath,amssymb}
\newcommand*\Laplace{\mathop{}\!\mathbin\bigtriangleup}
\newcommand{\Eq}[1]{Eq.~(\ref{#1})}

\newcommand{\App}[1]{appendix~\ref{#1}}

\newcommand{\ben}{\begin{enumerate}}
\newcommand{\een}{\end{enumerate}}
\newcommand{\bit}{\begin{itemize}}
\newcommand{\eit}{\end{itemize}}
\newcommand{\beq}{\begin{equation}}
\newcommand{\eeq}{\end{equation}}
\newcommand{\bsa}{\begin{subequations}\begin{eqnarray}}
\newcommand{\esa}{\end{eqnarray}\end{subequations}}
\newcommand{\bea}{\begin{eqnarray}}
\newcommand{\eea}{\end{eqnarray}}
\newcommand{\bean}{\begin{eqnarray*}}
\newcommand{\ean}{\end{eqnarray*}}

\newcommand{\Nf}{N_{\rm f}}
\newcommand{\Tr}{{\rm Tr}}

\begin{document}
\title{Optimising creation operators for charmonium spectroscopy on the lattice}
\begin{flushright}
\small{
WUB/22-02\\}
\vskip 0.3cm
\end{flushright}

\author{Francesco Knechtli}
\affiliation{Department of Physics, University of Wuppertal, Gau{\ss}strasse 20, 42119 Germany}
\author{Tomasz Korzec}
\affiliation{Department of Physics, University of Wuppertal, Gau{\ss}strasse 20, 42119 Germany}
\author{Michael Peardon}
\affiliation{School of Mathematics, Trinity College Dublin, Ireland}
\author{Juan Andrés Urrea-Niño}
\affiliation{Department of Physics, University of Wuppertal, Gau{\ss}strasse 20, 42119 Germany}

\begin{abstract}
\noindent Smearing the bare quantum fields in lattice calculations before applying composite hadron creation operators has a long record of substantially improving overlaps onto low-lying energy eigenstates. A technique called distillation which defines smearing for quark fields as a low-rank linear projection operator into a small vector space of smooth gauge-covariant fields has proven to be both effective and versatile in hadron spectroscopy calculations albeit with significant computational cost . In this paper, more general operators in this space of smooth fields are introduced and optimised, which enhances the performance of the method when tested on systems of heavy quark-anti-quark pairs close to the charmonium energy scale.
\end{abstract}
\maketitle
\section{Introduction}
\label{sec:intro}
\noindent In hadron spectroscopy, quark smearing is widely used to
improve the overlap between the creation operators applied and the desired 
state by
including quark wave-functions that are not point-like and instead have an
extended spatial distribution. A well-known
technique called \textit{distillation} was first proposed in
\cite{Peardon2009:Distillation} and has been widely used thanks to numerous
advantages. First, it provides a cut-off of the high eigenmodes of the 3D
lattice covariant Laplacian, which are exponentially suppressed by the Jacobi
smearing \cite{Allton1993:Jacobi}, by serving as a projector onto the subspace
spanned by low eigenmodes. Second, 
lattice quark propagation is represented via a
low-dimensional matrix, called the \textit{perambulator}. 
Since the lattice propagator is the 
inverse of a large-but-sparse matrix, practical calculations of correlation 
functions usually require point-to-all methods or stochastic estimations, and 
these are simplified in calculations involving smeared fields when written in 
terms of the perambulator.  The subspace is small enough to enable direct 
manipulation in many cases.  Third, 
since perambulators are independent of the creation operators, the 
inversion cost invested in their construction is fixed and correlation
functions of many arbitrary operators can be calculated efficiently.\\ \\

\noindent Unfortunately, this method also comes with disadvantages. Apart from the additional cost of calculating a number of eigenvectors of the Laplacian for each value of time in the lattice, the Dirac operator must be inverted a significant number of times using these eigenvectors. Namely, for a total $N_v$ of eigenvectors per time-slice of a lattice of temporal extent $N_t$ and $N_U$ gauge configurations, $4\times N_v \times N_t \times N_U$ inversions must be performed. The choice of $N_v$ is related to the level of smearing and, although commonly used values of $N_v$ are very small compared to the total number of eigenvectors the Laplacian has, this can still be of the order of $100$, which requires many inversions when large lattices and statistics are considered. Additionally, when the same level of smearing is desired on two different lattices, $N_v$ scales with the physical volume. This expense led to the introduction of stochastic estimation in the space of the Laplacian eigenvectors opposed to the actual calculation involving a large number of inversions \cite{Morningstar2011}. While greatly reducing the number of inversions required, this method uses approximations to the perambulators which depend on the estimation done and the dilution scheme of choice, therefore requiring a tuning of the parameters involved. The ultimate goal of this modification is to obtain results of comparable statistical quality to the exact case while keeping the number of inversions at a manageable level.\\ \\
\noindent In this work an improvement of the distillation method is proposed which still allows for the exact calculation of the perambulators, keeping the number of inversions needed untouched, yet taking advantage of a degree of freedom that neither standard distillation, nor its stochastic version, have exploited in detail before; the use of distillation profiles. The original formulation of the distillation operator allowed it to be expressed as
\beq
S[t] = V[t]J[t]V[t]^{\dagger}, \label{eq:S}
\eeq
where $V[t]$ has the eigenvectors on time slice $t$ as columns and the diagonal matrix $J[t]$ defines the restriction in the space spanned by $V[t]$. In Ref.~\cite{Peardon2009:Distillation}, the matrix $J[t]$ is set to unity. By construction, $S[t]$ has support only on time-slice $t$ so creation operators have explicit temporal locality. 
The spatial locality of the smeared fields has been explored \cite{Peardon2009:Distillation} and established empirically when $J=1$. It seems reasonable to expect using a smooth function of the eigenvalues to set diagonal entries in $J$, while all off-diagonal entries are kept fixed to zero will similarly result in a local smearing profile.
By exploring the effects of $J[t]\neq 1$ and generalising this freedom at the meson creation, it is expected that states with better overlap with the ones of interest can be constructed. The rest of this paper is organized as follows: Section \ref{sec:SmearProfOpti} explores the use of $J[t]\neq 1$ combined with a generalized eigenvalue problem to find the optimal $J[t]$. Section \ref{sec:CharmoniumSpectrum} displays the results of the charmonium spectrum in a model of QCD with $N_f = 2$ degenerate heavy quarks whose mass is about half of the value for the charm quark in nature. We find significant advantages of using the optimal distillation profiles when compared to standard distillation. Section \ref{sec:Conclusions} presents the conclusions of this work and future directions of the method are proposed here. Appendix \ref{app:OptimalGamma} shows how the optimal smearing profile can be equivalently expressed as an optimal creation operator for a meson.

\section{Optimal use of quark distillation profiles}
\label{sec:SmearProfOpti}

\noindent The starting point is the most general distillation operator $S[t]$ at a fixed time $t$ given by~\Eq{eq:S}.
$V[t]$ is a matrix with $4\times N_v$ columns constructed from eigenvectors of the covariant 3D Laplace operator $\Laplace[t]$. 
Since the operator does not act on Dirac components, $V[t]$ is a block identity in Dirac space and each block contains the first $N_v$ eigenvectors $v_i[t]$, sorted by magnitude of the corresponding eigenvalue $\lambda_i[t]$ in ascending order. A given column $V^{(i,\alpha)}[t]$ has entries given as
\beq
V^{(i,\alpha)}[t]_{\vec{x},t',\beta} = v_i[t]_{\vec{x}}\delta_{tt'}\delta_{\alpha \beta}.
\eeq 
$J[t]$ is a $4 N_v\times 4N_v$ matrix, also block diagonal in Dirac space. Its entries are given as 
\begin{equation}
J[t]_{\substack{ij\\\alpha \beta}} = g\left( \lambda_i[t] \right) \delta_{ij}\delta_{\alpha \beta}
\end{equation}
where $g\left( \lambda \right)$ is the \textit{quark distillation profile} and $\alpha, \beta = 0,...,3$ are Dirac indices. $S[t]$ can be written as
\beq
S[t]_{\substack{\vec{x},\vec{y}\\\alpha \beta}} = \delta_{\alpha \beta}\sum_{i=1}^{N_v} v_i[t]_{\vec{x}} g\left( \lambda_i[t] \right) v_i[t]^{\dagger}_{\vec{y}}
\eeq
and the role of $J[t]$, and therefore $g(\lambda)$, becomes apparent; it serves as the modulation of the contribution from each of the eigenvectors. The choice $g\left( \lambda \right) = 1$ turns $S[t]$ into an orthogonal projector onto the range of $V[t]$ and is the most commonly used one\cite{Liu2012, Cheung2016, Dudek2010, Dudek2013,zhang2021glueball, zhang2021annihilation}. 
In \cite{Dudek2010} also a single gaussian profile was tested. It should be noted that, while remarkable results are obtained with this choice, 
the exponential suppression caused by the Jacobi smearing, the inspiration for distillation, indicates that it is not a requirement that all $v_i[t]$ contribute with the same weight that they have in the original quark field when constructing an optimally smeared quark field. With this in mind, a step to improve over the standard distillation projector commonly used consists on taking advantage of the freedom of choice of $g(\lambda)$ to build an optimal meson operator based on distilled quark fields. The strategy to follow consists of fixing an operator and building a basis of size $N_{B}$ with different $g_k\left( \lambda \right)$, $k=1,...,N_B$, from which a variational formulation \cite{Blossier2009:GEVP} helps to extract the optimal meson operator.

\subsection{Quark distillation profile in meson two-point correlations}
\noindent Following the procedure in \cite{Peardon2009:Distillation}, consider the meson operator 
\beq
\mathcal{O}(\vec{x},t) = \bar{q}(\vec{x},t) \Omega \Gamma q(\vec{x},t),
\label{eqn:MesonOperator}
\eeq
where $q$ is a doublet of mass-degenerate quarks $q=(q_1,q_2)$, $\Omega \in \{{\mathbb{I}, \tau_3}\}$ is a $2\times2$ flavour matrix and $\Gamma$ is an operator that fixes the symmetry channel. The quark fields $q_i$ are replaced in Eq. \eqref{eqn:MesonOperator} by their distilled counterparts $\tilde{q}_i$ defined as
\beq
\tilde{q}_i(\vec{x},t)_{\alpha} = \sum_{\vec{y},\beta} S[t]_{\substack{\vec{x}, \vec{y}\\\alpha \beta}} q_i(\vec{y},t)_{\beta},
\eeq
to build the distilled meson operator 
\beq
\tilde{\mathcal{O}}(\vec{x},t) = \bar{\tilde{q}}(\vec{x},t) \Omega \Gamma \tilde{q}(\vec{x},t).
\eeq
This operator can then be projected to zero spatial momentum using
\beq
\hat{\mathcal{O}}(t) = \sum_{\vec{x}} \tilde{\mathcal{O}}(\vec{x},t),
\eeq
and the two-point meson correlation function $C(t',t)$ can be written as
\beq
C(t',t) = \left \langle \hat{\mathcal{O}}(t') \bar{\hat{\mathcal{O}}}(t) \right \rangle_{U,F},
\eeq
where the subscripts $U$ and $F$ denote the averaging over gauge and fermion fields. The integration over the fermion fields can be done analytically, yielding \footnote{when $\hat{\mathcal{O}}(t)$ has non-zero vacuum expectation value the replacement $\hat{\mathcal{O}}(t) \rightarrow \hat{\mathcal{O}}(t) - \left \langle \hat{\mathcal{O}}(t) \right \rangle_{F,U}$ is done.}
\begin{align}
\begin{split}
C(t',t) &= -2\left \langle \Tr\left( \Phi^{(\Gamma)}[t']\tau[t',t]\bar{\Phi}^{(\Gamma)}[t]\tau[t,t'] \right)\right \rangle_{U}\\
&+ \Tr\left( \Omega \right)^2 \left \langle \Tr\left( \bar{\Phi}^{(\Gamma)}[t]\tau[t,t]\right) \Tr\left( \Phi^{(\Gamma)}[t']\tau[t',t'] \right) \right \rangle_{U},\\
\end{split}
\label{eqn:Meson2Point}
\end{align}
where $\Tr\left( \Omega \right)$ can either be $2$ for $\Omega = \mathbb{I}$, the iso-scalar case, or $0$ for $\Omega = \tau_3$, the iso-vector case. The \textit{modulated elementals} $\Phi^{(\Gamma)}[t]$ have entries
\beq
\Phi^{(\Gamma)}_{\substack{\alpha \beta\\ij}}[t] = \mathcal{H}_{\alpha \beta} g\left( \lambda_i[t] \right)^{*}v_i[t]^{\dagger}\mathcal{D}[t] v_j[t] g\left( \lambda_j[t] \right)
\eeq
and the \textit{perambulators} $\tau[t,t']$ have entries
\beq
\tau_{\substack{\alpha \beta\\ij}}[t',t] = v_{i}[t']^{\dagger}D^{-1}_{\substack{t't\\\alpha \beta}}v_{j}[t],
\eeq
where the quark propagator $D^{-1}$ includes the dependence on the mass of the degenerate quarks. For clarity the operator $\Gamma[t]$ is explicitly separated into an operator $\mathcal{H}$ that acts on spin space and an operator $\mathcal{D}[t]$ that acts on color and coordinate space, the latter being the one that can act on the eigenvectors. These definitions can be written in matrix form as 
\begin{align}
\begin{split}
\Phi^{(\Gamma)}[t] &= J[t]^{\dagger}V[t]^{\dagger} \Gamma[t] V[t] J[t]\\ &=\mathcal{H} J[t]^{\dagger} V[t]^{\dagger}\mathcal{D}[t]V[t]J[t]
\end{split}
\label{eqn:ElementalDefinition}
\end{align}
and
\beq
\tau [t',t] = V[t]^{\dagger}D^{-1}V[t],
\eeq
It can be seen that the modulated elementals contain a factor independent of the choice of $\left( \lambda \right)$; the \textit{elementals} of $\Gamma = \mathcal{H}\mathcal{D}$
\beq
\Lambda^{(\Gamma)}_{\substack{\alpha \beta\\ij}}[t] = \mathcal{H}_{\alpha \beta}v_i[t]^{\dagger}\mathcal{D}[t]v_j[t]. 
\label{eqn:OriginalElemental}
\eeq
There is a two-fold advantage in defining and calculating these elementals. On one hand, once they are calculated it is possible to introduce the factor $g\left( \lambda_i[t] \right)^{*}g\left( \lambda_j[t] \right)$ for different choices of distillation profiles. This represents a notable gain as expensive operations, such as the application of an arbitrary number of lattice derivatives, have to be performed only once while the factor including the distillation profile defines the different $N_B$ operators. On the other hand, the elementals of operators that contain linear combinations of an arbitrary number of lattice derivatives are a linear combination of derivative elementals defined as
\beq
\Lambda_{ij}^{[m...n...l]}[t] = v_{i}[t]^{\dagger} \nabla_m[t] ...\nabla_n[t] ... \nabla_l[t] v_j[t],
\label{eqn:derivativeElemental}
\eeq  
where $\nabla_i[t]$ is the symmetric gauge covariant lattice derivative in direction $i$. These derivative elementals need to be calculated once and can then be combined in different ways depending on the operator and scaled according to the chosen distillation profile. Additionally, the number of derivatives involved in these derivative elementals leads to further simplifications. For the case where there are no derivatives, no elemental needs to be calculated due to the 
orthogonality of the eigenvectors for a fixed time, i.e $v_i[t]^{\dagger} v_j[t] = \delta_{ij}$. The elemental in such a case is given by
\beq
\Phi^{(\Gamma)}_{\substack{\alpha \beta\\ij}}[t] = \mathcal{H}_{\alpha \beta} \delta_{ij}g\left( \lambda_i[t] \right)^{*}g\left( \lambda_j[t] \right).
\eeq
For the case with a single derivative, the elemental is given by
\beq
\Lambda_{ij}^{[m]}[t] = v_{i}[t]^{\dagger} \nabla_m[t] v_j[t]
\eeq
and satisfies $\Lambda_{ij}^{[m]}[t] = - \Lambda_{ji}^{[m]}[t]^{*}$, i.e it is anti-hermitian. This means that for a fixed time $t$ only $\frac{1}{2}N_v(N_v + 1)$ entries must be calculated, instead of $N_v^2$, via only $N_v$ lattice derivatives and $\frac{1}{2}N_v(N_v + 1)$ inner products. The case with two derivatives yields a derivative elemental
\beq
\Lambda_{ij}^{[mn]}[t] = v_{i}[t]^{\dagger} \nabla_m[t] \nabla_n[t] v_j[t]
\eeq
that satisfies $\Lambda_{ij}^{[mn]}[t] = \Lambda_{ji}^{[nm]}[t]^{*}$, i.e $\Lambda^{[mn]}[t] = \Lambda^{[nm]}[t]^{\dagger}$. Due to the fact that lattice derivatives in different directions do not commute, all the entries of $\Lambda^{[mn]}[t]$ must be calculated for fixed $t$, $m$ and $n$ if $m \neq n$. However, the aforementioned property means that only three pairs $[mn]$ satisfying $m\neq n$ out of the possible 6 need to be calculated. For the case $m=n$ this same property dictates that only $\frac{1}{2}N_v (N_v + 1)$ entries must be calculated. The case with three or more derivatives is not treated here yet similar relationships between the derivative elementals and their hermitian conjugates are expected.

\subsection{Determination of the optimal meson distillation profile}
\noindent We consider a variational basis of meson operators $\mathcal{O}_i$ with different quark smearing profiles $g_i(\lambda)$ in a fixed symmetry channel $\Gamma = \mathcal{H} \mathcal{D}$. From such a basis, operators can be constructed that have the best overlap with a given energy eigenstate in this channel.
The starting point is the correlation matrix
\beq
C_{ij}(t) = \left\langle \mathcal{O}_i(t)\bar{\mathcal{O}}_j(0) \right\rangle_{F,U}.
\eeq
For a fixed value $t_G$ the generalized eigenvalue problem (GEVP) given by
\beq
C(t)w_e(t,t_G) = \rho_e(t,t_G)  C(t_G)w_e(t,t_G)
\label{eqn:GEVP}
\eeq
is solved for $t > t_G$ and $e=0,...,N_B-1$ for the generalized eigenvalues $\rho_e(t,t_G)$ and eigenvectors $w_e(t,t_G)$.
As shown in \cite{Luscher1990, Blossier2009:GEVP, Irges2007}, the generalized eigenvalues $\rho_e(t,t_G)$, when periodic boundary conditions in the gauge field are used, behave in the large t limit as
\beq
\rho_e(t,t_G) = 2c_{e} e^{-\frac{T}{2}m_e}\cosh \left( \left( \frac{T}{2} - t\right) m_e \right),
\label{eqn:Eigenvalues}
\eeq
where $c_e$ is a constant, $m_e$ is the mass of the state of interest and $T$ is the temporal extent of the lattice in physical units. Ordering the generalized eigenvalues from largest to smallest, $e=0$ 
corresponds to the ground state and $e=1,...,N_B-1$ to the further excited states. The effective mass $m_e(t)$ can be extracted for $t=t_G + a,...,\frac{T}{2} - a$ via a root finding method as shown in \cite{Irges2007}. This GEVP formulation allows access to excited states but it may suffer from numerical instabilities when many similar operators are considered for the construction of $C(t)$. This issue can be attenuated by following the strategy presented in \cite{Balog1999, Niedermayer2001}. 
At a chosen time separation $t_S$ a singular value decomposition of $C(t_S)$ is performed.
The first $N_S$ singular vectors corresponding to the largest $N_S$ singular values are used to define a pruned correlation matrix
\beq
C_{S}(t)_{ij} = u_{i}^{\dagger} C(t)u_j,
\label{eqn:PrunedCorrelation}
\eeq
where $u_i$ are the selected singular vectors for $i=0,...,N_S-1$. 
The pruned correlation matrix is better conditioned by construction, and less prone to the numerical instabilities mentioned above. 
Generalized eigenvalues obtained from solving the GEVP of $C_{S}(t)$ are then used to extract the effective mass \cite{Irges2007}. Another use of the GEVP formulation is the fact that the entries of the generalized eigenvectors 
give the coefficients of an optimal linear combination of $\mathcal{O}_i$ that achieves the best possible overlap with the energy state of interest. The first step to find the explicit form of the optimal operators is to transform the original basis into the basis of the pruned matrix $C_{S}(t)$. The original meson profile basis is 
\beq
f_k(\lambda_i, \lambda_j) = g_{k}(\lambda_i)^{*}g_{k}(\lambda_j), 
\eeq
for $k=0,...,N_B -1$ and the pruned profile basis is constructed from the singular vectors $u_j$ as
\beq
f^{(p,\Gamma)}_{n}(\lambda_i,\lambda_j) = \sum_{m=0}^{N_B - 1} u_{n,m} f_m(\lambda_i, \lambda_j),
\label{eqn:PrunedProfile}
\eeq
with $n=0,...,N_S - 1$ and where $u_{n,m}$ denotes the $m$-th entry of the $n$-th singular vector. Note that at this point a dependence on $\Gamma$ is introduced via the matrix $C_{S}(t)$. The second and final step is to linearly combine the pruned profiles using the generalized eigenvectors $w^{(\Gamma, e)}$ which yields
\beq
\tilde{f}^{(\Gamma,e)}(\lambda_i, \lambda_j) = \sum_{m=0}^{N_S - 1} w_{m}^{(\Gamma, e)} f_{m}^{(p, \Gamma)}(\lambda_i, \lambda_j).
\label{eqn:OptimalMesonProfile}
\eeq
Here $w_{k}^{(\Gamma, e)}$ denotes the $k$-th entry of the $e$-th  generalized eigenvector in the channel $\Gamma$ and the $t,t_G$ dependence is suppressed.
With the optimal meson profile $\tilde{f}^{(\Gamma,e)}(\lambda_i, \lambda_j)$ the optimal meson elemental can be defined, now including the time dependence, as
\beq
\Phi^{(\Gamma, e)}_{\substack{\alpha \beta\\ij}}[t] = \tilde{f}^{(\Gamma,e)}(\lambda_i[t], \lambda_j[t]) \Phi^{(\Gamma)}_{\substack{\alpha \beta\\ij}}[t],
\eeq
or in matrix form as
\beq
\Phi^{(\Gamma, e)}[t] = \Phi^{(\Gamma)}[t] \circledast F^{(\Gamma, e)}[t],
\label{eqn:OptimalElementalMatrix}
\eeq
\noindent where $\circledast$ denotes the element-wise product and $F^{(\Gamma, e)}[t]$ is given by
\beq
F^{(\Gamma, e)}[t]_{ij} = \tilde{f}^{\left( \Gamma, e\right)}\left( \lambda_i[t], \lambda_j[t] \right).
\eeq
\noindent Eq. \eqref{eqn:OptimalElementalMatrix} shows the main difference between distillation using a quark profile and a meson profile: in the first case the profile is inserted manually at quark level via the distillation operator $V[t]J[t]V[t]^{\dagger}$ while in the second case the profile is inserted manually at meson level in the elemental via $F[t]$ yet at quark level just a projection is done via $V[t]V[t]^{\dagger}$. This meson distillation profile can be thought of as the meson's wavefunction in distillation space and modulates the coupling between $v_i[t]$ and $\mathcal{D}[t]v_j[t]$ based on the corresponding eigenvalues, spin structure $\mathcal{H}$ and the energy level $e$. The resulting optimal meson distillation profile has the special feature of not being separable in general for $N_B > 1$, i.e there is no quark smearing profile $g_q^{\left( \Gamma, e\right)}(\lambda)$ such that 
\beq
\tilde{f}^{\left( \Gamma, e\right)}\left( \lambda_i[t], \lambda_j[t] \right) =  g_q^{\left( \Gamma, e\right)}\left(  \lambda_i[t]\right)^{*} g_q^{\left( \Gamma, e\right)}\left( \lambda_j[t] \right)
\eeq
and therefore it is not possible to define an optimal quark smearing profile $J_q^{\left( \Gamma, e \right)}[t]$ with entries given by
\beq
J_q^{\left(\Gamma, e \right)}[t]_{ij} = \delta_{ij} g_q^{\left( \Gamma, e \right)}\left( \lambda_i[t] \right).
\eeq
With this in mind it is more convenient to not attempt to translate these meson profiles to quark ones but rather keep the analysis at meson level. One can shift the attention to the $\Gamma$ structure that is in the meson operator and define an optimal $\tilde{\Gamma}_D[t]= V[t] \tilde{\Phi}[t] V[t]^{\dagger}$ structure which couples the quark fields, includes in its definition the optimal meson profile of the desired energy state of the channel of interest and yields the same correlation function as the one obtained from the GEVP described above (See \App{app:OptimalGamma}).

\section{Charmonium spectrum}
\label{sec:CharmoniumSpectrum}
\begin{figure}[H]
\centering
\includegraphics[width=0.75\linewidth]{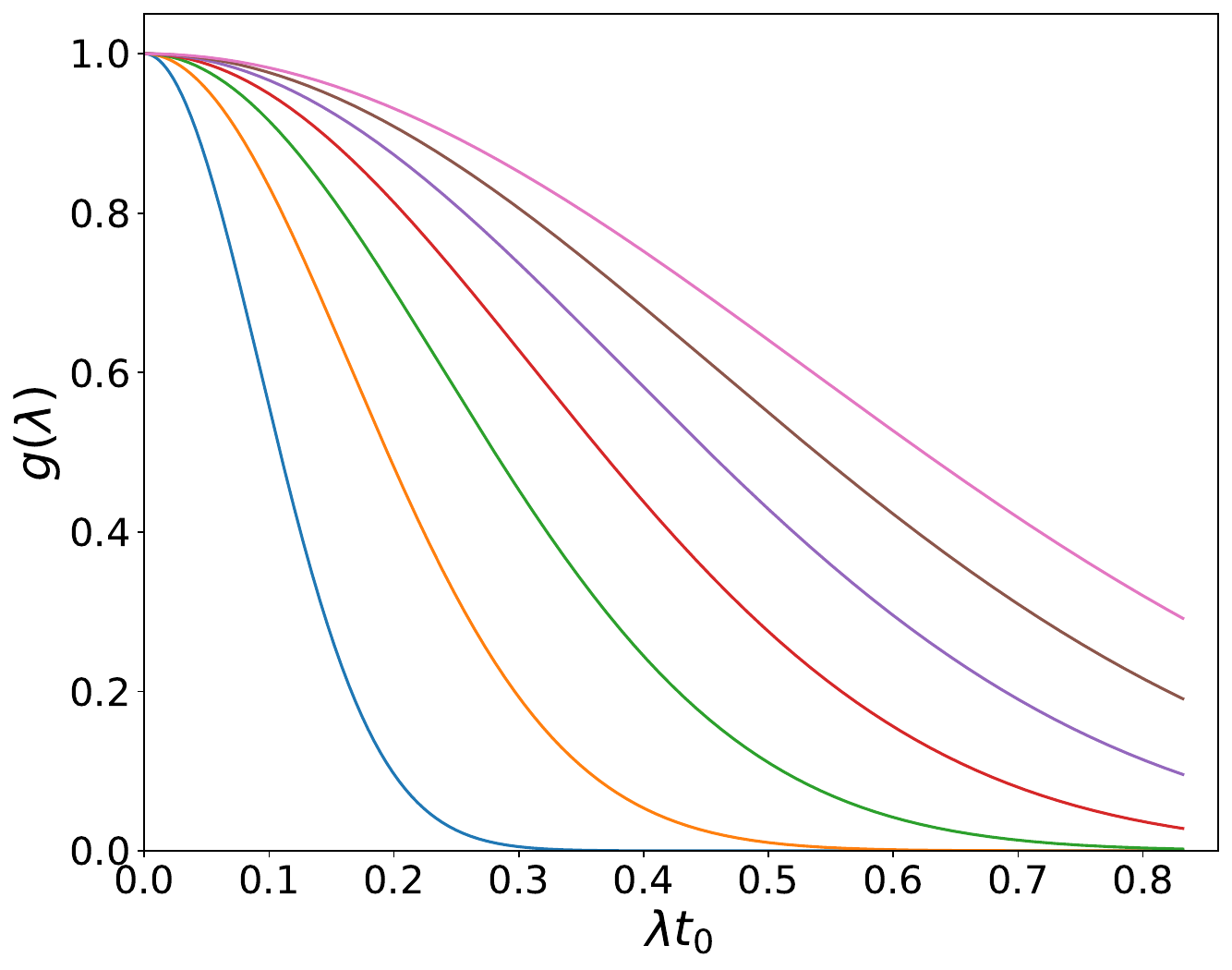}
\caption{Input quark distillation profiles used for the GEVP.}
\label{fig:ProfileBasis}
\end{figure}

\noindent All the calculations are performed on an ensemble with a $48\times 24^3$ lattice generated with two dynamical non-perturbatively O($a$) improved Wilson quarks \cite{Jansen:1998mx} with a mass equal to half of the physical charm. We use periodic boundary conditions except for anti-periodic boundary conditions for the fermions in the temporal direction. The bare gauge coupling is $g_0^2=6/5.3$ and the hopping paramater is $\kappa=0.13270$. The lattice spacing is a=0.0658(10) fm \cite{Fritzsch:2012wq,Cali:2019enm} and the flow scale \cite{Luscher:2010iy} $\frac{t_0}{a^2} = 1.8486(7)$. A total of $N_v = 200$ eigenvectors of the 3D covariant Laplacian are used in each time-slice of the lattice. These are calculated with a Chebyshev accelerated \cite{SorensenChebyshev} Thick-Restart Lanczos algorithm \cite{Wu2000} with periodic reorthogonalization \cite{GrcarPeriodic, Simon1984} in the first run and full reorthogonalization in the restarts. The Chebyshev acceleration uses a carefully chosen Chebyshev polynomial as a spectral filter which shifts and spreads out the segment of the spectrum one is interested in, therefore accelerating the convergence of the Lanczos algorithm. The periodic reorthogonalization reduces the computational cost of the algorithm while still keeping its precision. In this work the lower end of the spectrum is subjected to the filtering, however one could choose the upper end of the spectrum as well since there is a one-to-one relationship between the eigenvalues and eigenvectors of the 3D covariant Laplacian at both sides of the spectrum \cite{Greensite2005}. A total of 20 3D APE smearing \cite{Albanese1987} steps with $\alpha_{APE} = 0.5$ are applied on each gauge field before the eigenvector calculation so as to smooth the link variables that enter the Laplacian operator. These parameters were found to yield the best results in a study of the static potential on the same ensemble. No smearing is applied to the gauge field used for the derivatives $\nabla_i$. All calculations are perfomed by a code based on QCDlib, a library written by us in C+MPI that facilitates massively parallel QCD calculations. The inversions of the Dirac operator are performed by calling the package openQCD\footnote{http://luscher.web.cern.ch/luscher/openQCD/} \cite{Luscher2013}, namely a deflated SAP GCR solver with improvements based on the two-grid method of \cite{Frommer:2013fsa}. The error analysis in this work is done using the $\Gamma$ method \cite{Wolff2004,Schaefer2011}. The $N_B$ quark distillation profiles $g_k(\lambda)$ used are given by
\beq
g_k(\lambda) = \exp\left\{-\dfrac{\lambda^2}{2\sigma_{k}^2}\right\},
\eeq
where the $\sigma_k$ define the width of the gaussians. This basis was chosen due to two reasons. First of all, each $g_k(\lambda)$ enforces a suppression of large eigenvalues, just as Jacobi smearing does, and follows distillation's intuition that small eigenvalues contribute more than large ones. Second of all, different basis functions, such as the monomials $\lambda^{k}$, were tried with low statistics and the gaussian basis resulted in the most numerically stable GEVPs. The values $\sigma_k$ are chosen such that a wide range of widths is covered. $N_B = 7$ is fixed and the widths are equally spaced between $\sigma_1=0.0924/t_0$ and $\sigma_7=0.7949/t_0$. This choice means that the broadest profile still allows for a non-negligible contribution from the 200-th eigenvalue while the thinnest has already majorly suppressed it. This way the entire range of the $200$ eigenvalues is covered and can be either enhanced or suppressed. The resulting quark distillation profiles can be seen in Fig. \ref{fig:ProfileBasis}. The pruning of the correlation matrix mentioned in the previous section is done by taking $N_S = 4$ singular vectors for all the operators studied in this work at a time $t_S = 3a$ which also corresponds to the value of $t_G$ which is fixed for the GEVP formulation.

\begin{table}[H]
\centering
\begin{tabular}{||c c c||} 
 \hline
 $J^{PC}$ & $\Gamma$ & Particle\\ [0.5ex] 
 \hline\hline
 $0^{-+}$ & $\gamma_5$ & $\eta_c$, $\eta_c (2S)$,...\\ 
 & $\gamma_0 \gamma_5 \gamma_i \nabla_i$ & \\
 & $\gamma_i \mathbb{B}_i$ & \\
 \hline
 $1^{--}$ & $\gamma_i$ & $J/\Psi$, $\Psi(2S)$,...\\
 & $\nabla_i$ & \\
 & $\gamma_5 \mathbb{B}_i$ & \\
 \hline
 $0^{++}$ & $\mathbb{I}$ & $\chi_{c0}$,...\\
          & $\gamma_i \nabla_i $ & \\
 \hline
 $1^{++}$ & $\gamma_5 \gamma_i$ & $\chi_{c1}$,...\\
          & $\epsilon_{ijk}\gamma_j \nabla_k$ & \\
 \hline
 $1^{+-}$ & $\epsilon _{ijk} \gamma_j \gamma_k$ & $h_c$,...\\ 
  & $\gamma_5 \nabla_i$ & \\
 \hline
 $2^{++}$ & $|\epsilon_{ijk}|\gamma_j \nabla_k$ $\left( T_2\right)$ & $\chi_{c2}$,...\\ 
  & $\mathbb{Q}_{ijk}\gamma_j \nabla_k$ $\left( E \right)$ & \\
 \hline
 $1^{-+}$ & $\gamma_0 \nabla_i$ & -\\
  & $\epsilon_{ijk}\gamma_j \mathbb{B}_k$ & \\ [1ex]
 \hline
\end{tabular}
\caption{$J^{PC}$ channels studied with their corresponding 
creation operators.
}
\label{table:OperatorList}
\end{table}

\noindent For the charmonium spectrum the $J^{PC}$ of interest are $0^{-+}$, $1^{--}$, $0^{++}$, $1^{++}$, $1^{+-}$, $2^{++}$ and $1^{-+}$, with the last one being a spin-exotic quantum number. Table \ref{table:OperatorList} shows the operators that transform according to the irreducible representations of the cubic group that contain a contribution from the $J^{PC}$ of interest. Local operators are used due to their ease of computation. Derivative based operators, here taken from \cite{Dudek2008Operators}, are used to sample different spatial structures \cite{Peardon2009:Distillation, Dudek2010, Burch2009, Gattringer2008}, explore $J^{PC}$ not available via local operators \cite{Dudek2010, Gattringer2008, Dudek2008Operators} and also ones that might include gluonic degrees of freedom \cite{Dudek2008Operators, Liu2012, Dudek2009}. 
One example is the operator $\epsilon_{ijk}\gamma_j \mathbb{B}_k$, where $\mathbb{B}_i = \epsilon_{ijk}\nabla_j \nabla_k$ is proportional to $\epsilon_{ijk} F^{jk}$ in the continuum and therefore explicitly contains the field-strength tensor. An important difference between local and derivative-based operators should be mentioned at this point: while for local operators the elementals are diagonal in distillation space, for derivative-based operators this is not the case. This is due to the fact that the latter act on coordinate/color space and therefore act directly on the eigenvectors that are used. In this sense the elementals retain the information of the original derivative-based operator that exists only in the span of these eigenvectors, just as the perambulators do with respect to the inverse of the Dirac operator. All operators were measured on 4080 gauge configurations. Some of the following results were already presented in \cite{UrreaProc}.

\subsection{Using quark-connected correlations}
\begin{figure}[H]
\centering
\includegraphics[width=0.9\linewidth]{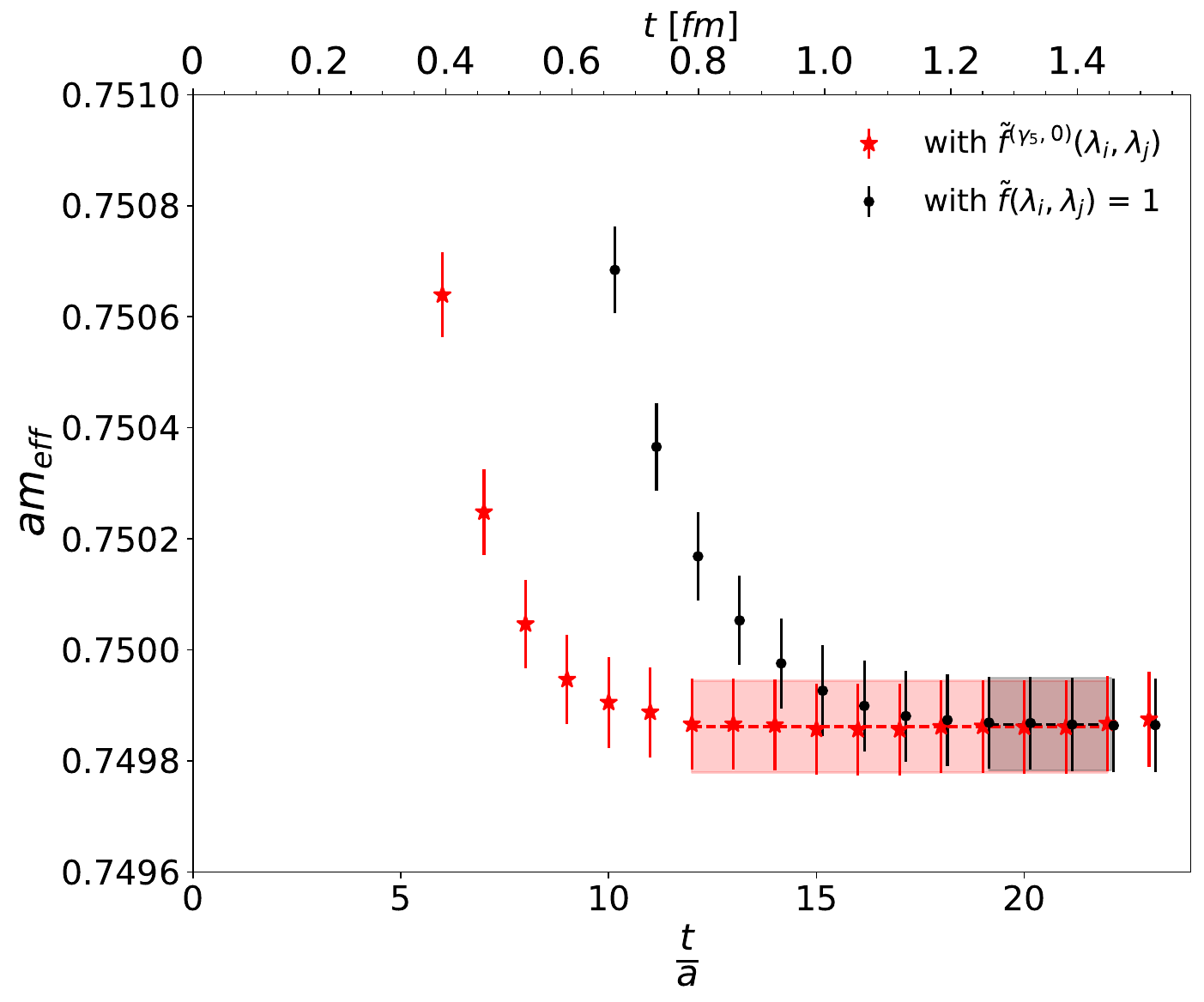}
\caption{Effective mass of ground state for $\Gamma = \gamma_5$ using both constant ($\tilde{f}(\lambda_i, \lambda_j) = 1$) and optimal $\tilde{f}^{(\gamma_5,0)}(\lambda_i, \lambda_j)$ meson profiles.}
\label{fig:g5Ground}
\end{figure}

\begin{figure}[H]
\centering
\includegraphics[width=0.45\textwidth]{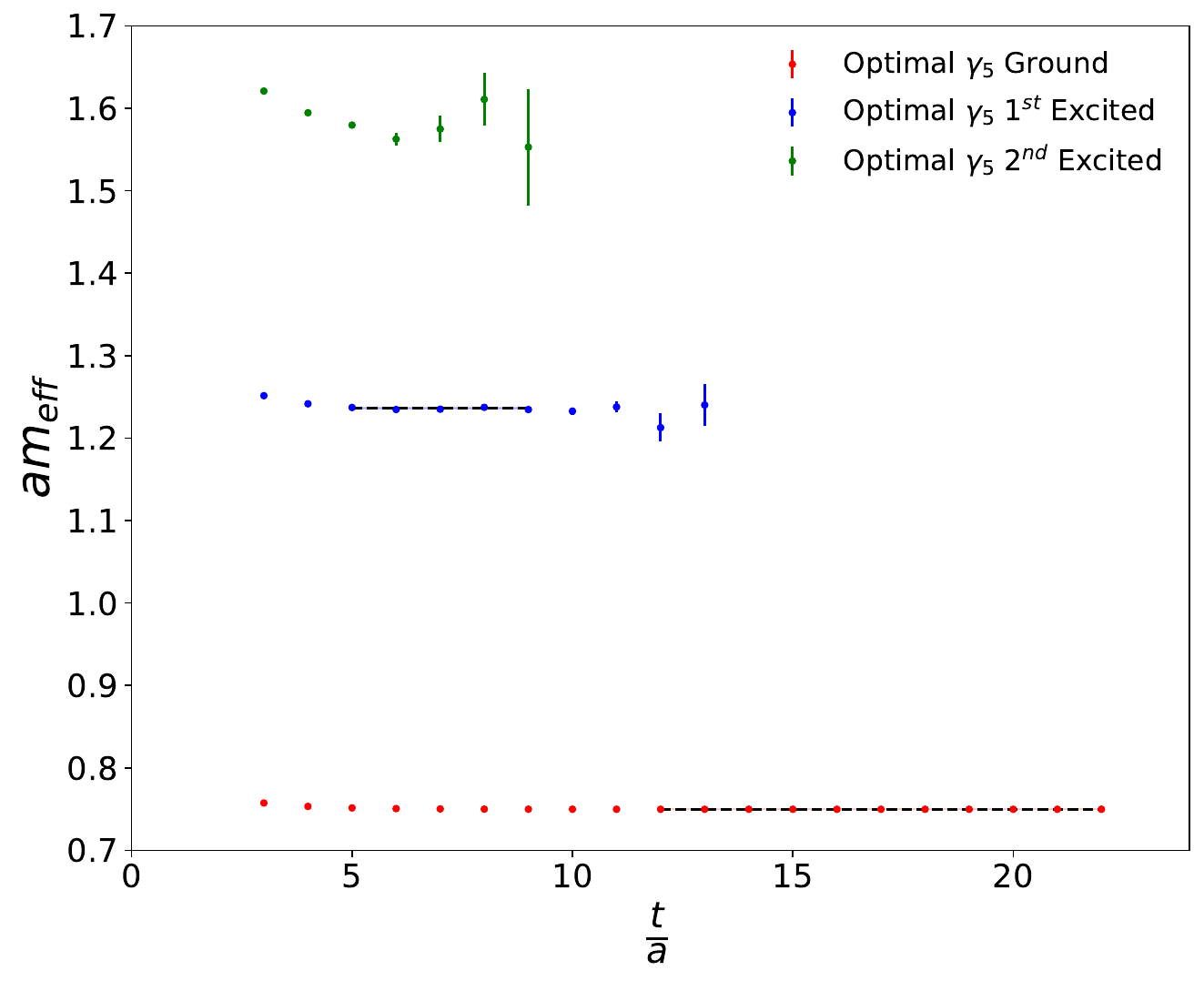}
\caption{Effective mass of ground, first- and second-excited
states for $\Gamma = \gamma_5$ using the optimal $\tilde{f}^{(\gamma_5,e)}(\lambda_i,\lambda_j)$ $(e=0,1,2)$ meson profiles generated by the GEVP. Single hyperbolic cosine fit results are included for the ground and first-excited states.}
\label{fig:G5_AllStates}
\end{figure}

\begin{figure}[H]
\centering
\includegraphics[width=0.9\linewidth]{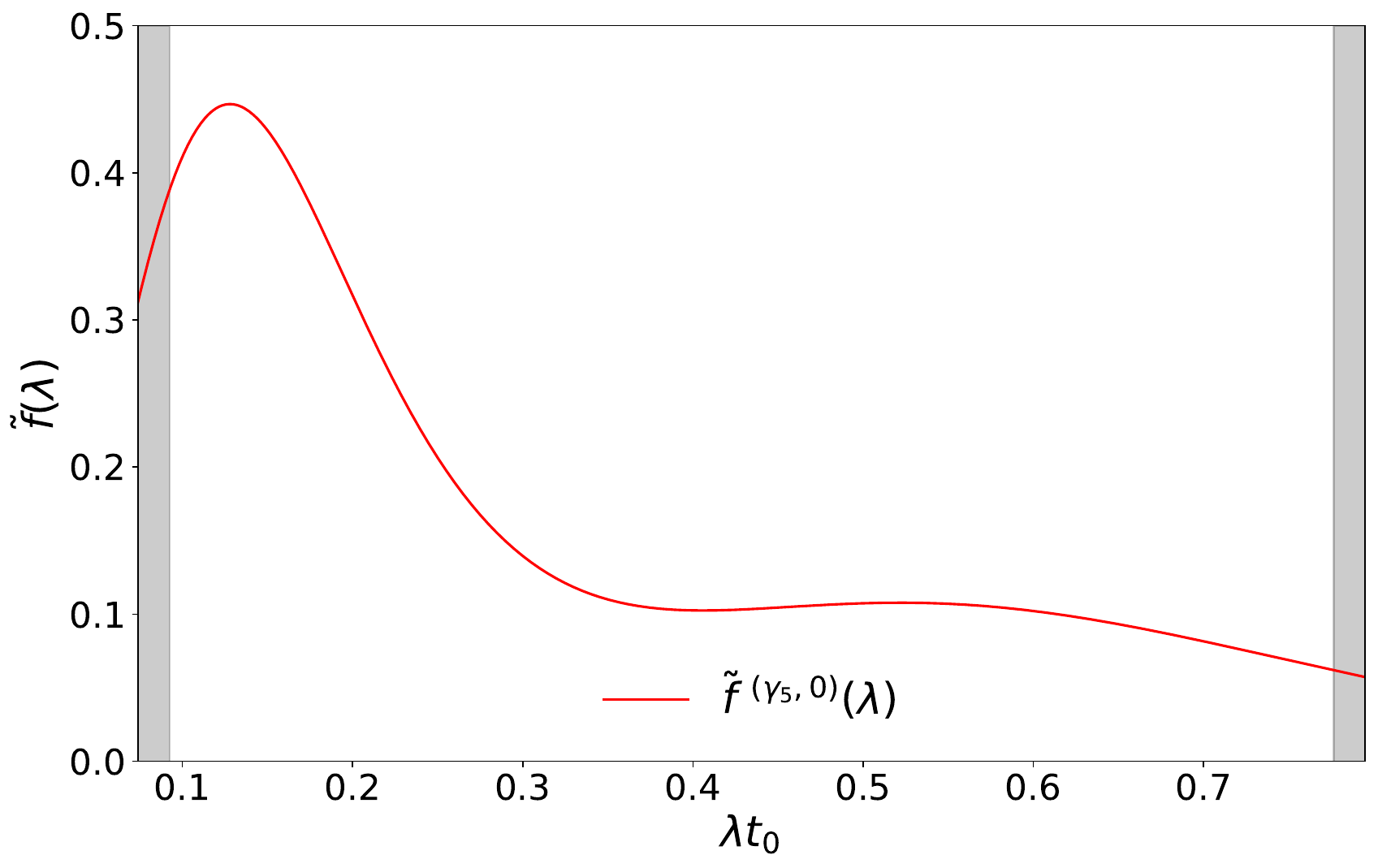}
\caption{Optimal meson distillation profile for the ground state of the iso-vector $\Gamma = \gamma_5$ operator, $\tilde{f}^{(\gamma_5,0)}(\lambda_i, \lambda_j)$, evaluated at $\lambda_i = \lambda_j = \lambda$}.
\label{fig:g5Ground_Profile}
\end{figure}

\noindent The first results to be presented correspond to the iso-vector spectrum and are obtained when $\Omega = \tau_3$ in Eq. \eqref{eqn:MesonOperator}. The results for different $J^{PC}$ using local operators will be presented first, followed by derivative-based operators. The first channel of interest to be studied is $J^{PC} = 0^{-+}$ and Fig. \ref{fig:g5Ground} displays a zoomed-in plot of the effective masses obtained for the ground state of the local operator $\gamma_5$ built using both standard distillation and the optimal meson distillation profile from the GEVP analysis. The use of the optimal meson profile results in a notable suppression of excited state contamination and therefore an earlier mass plateau. Fig. \ref{fig:G5_AllStates} shows the effective masses of the lowest three states generated by the GEVP using the basis of operators formed from the seven smearing profiles in the bilinear $\Gamma = \gamma_5$ alone. Clear signals for the first- and second-excited states are resolved using this single spin bilinear in with multiple meson profiles. Fig. \ref{fig:G5_AllStates} includes the result of a fit of the correlation functions to a single hyperbolic cosine. Given this improvement, the form of the optimal meson profile that corresponds to this ground state is of interest. It can be easily obtained from the formulation presented in the previous section. Since the operator is point-like, the elemental is diagonal in distillation space and only the values of $\tilde{f}^{(\gamma_5,e)}(\lambda_i, \lambda_j)$ with $\lambda_i = \lambda_j$ are relevant. As a result, this profile can be plotted as a function of only one variable. Fig. \ref{fig:g5Ground_Profile} displays this optimal meson profile. 
It is only known in the interval in which the eigenvalues are contained, which in the plot corresponds to the region between the two gray shaded areas. The error of this profile, which is also a function of the eigenvalues, has values of order $10^{-3}$ so the corresponding errorbars are omitted since they would not be clearly visible at the scale of the plot. Two important pieces of information can be taken from Fig. \ref{fig:g5Ground_Profile}. First, this optimal profile is not a constant, as standard distillation enforces, which implies that, apart from the truncation of Laplacian eigenmodes, a modulated filtering of these remaining eigenmodes can significantly improve the results. Second, this filter, given here by the meson distillation profile, is not arbitrary but rather obeys one of the ideas behind distillation; low eigenmodes contribute considerably more than higher ones and therefore the weights they have in the entries of the elemental must enhance them accordingly.

\newpage

\begin{figure}[H]
\centering
\includegraphics[width=0.9\linewidth]{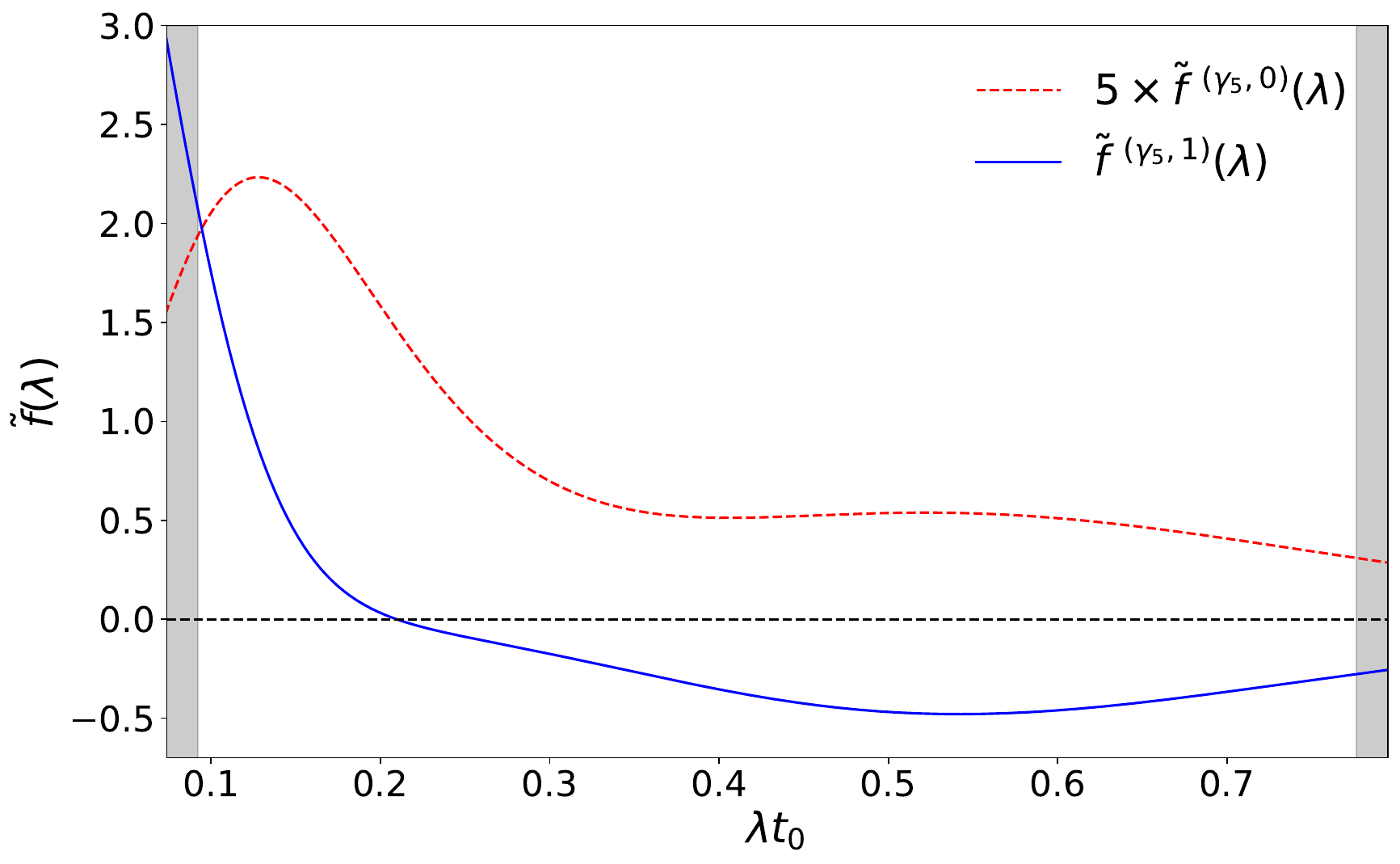}
\caption{Optimal meson distillation profile for the first excited state of the $\Gamma = \gamma_5$ operator. The scaled profile corresponding to the ground state is also included for comparison.}
\label{fig:g5First_Profile}
\end{figure}

\noindent It is interesting to see whether such a picture emerges  for the excited states as well. Therefore the first excited state for this channel is also analyzed. The optimal meson distillation profile for that state is shown in Fig. \ref{fig:g5First_Profile} together with a scaled version of the profile for the ground state for comparison. The errorbars for this excited state profile are also omitted since the corresponding errors are at the percent level. Similar observations as for the ground state can be made regarding the importance of low eigenmodes compared to high ones and the need for further filtering. However, and unlike the case of the ground state, the profile for this first excited state displays a node. 
The magnitude of the profile has a second local maximum at larger eigenvalues, indicating that for excited states higher modes are more importand than for the ground state. Moreover the suppression of the highest modes is not as pronounced as it was for the ground state. Profiles like these can serve as a guide for the choice of a reasonable number eigenmodes to work with.

\begin{figure}[H]
\centering
\includegraphics[width=0.9\linewidth]{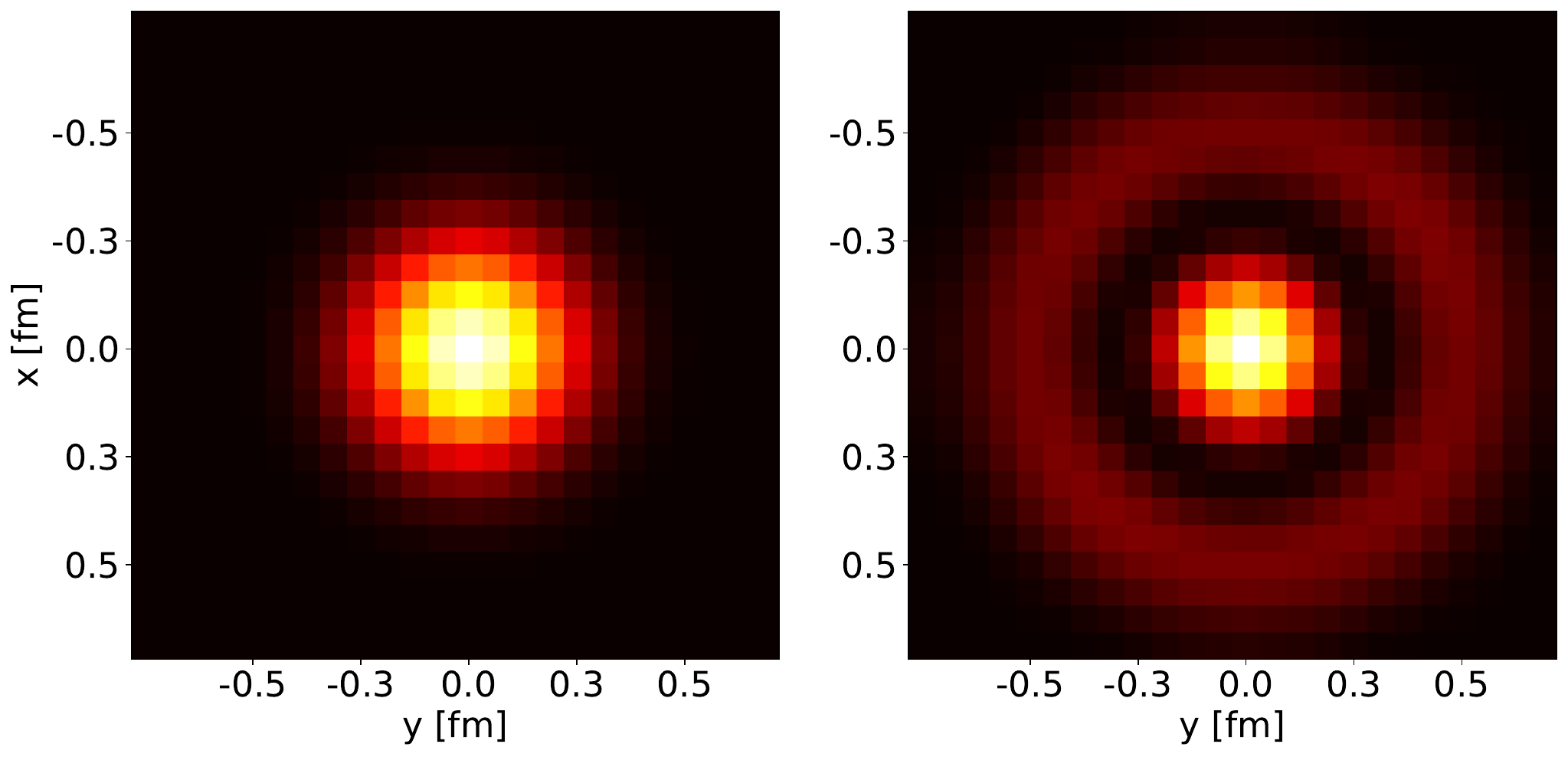}
\caption{Spatial distributions $\Psi^{(\gamma_5,0)}(\vec{x})$ and $\Psi^{(\gamma_5,1)}(\vec{x})$. The coloring changes from dark (zero) to light as the values of the profile increase.}
\label{fig:SpatialProfiles_G5}
\end{figure}

\noindent The second excited state was also analyzed and although the effective mass data becomes rather noisy it is still possible to extract the optimal meson distillation profile for this state. In this case it displays two nodes, and even higher modes contribute significantly.
In addition to the profiles in ``eigenvalue-space'', it is interesting to visualize the corresponding spatial profiles of the optimal $\tilde{\Gamma}[t]$ for both ground and first excited state. They can be constructed from the optimal elementals and their  explicit form is
\begin{equation}
\tilde{\Gamma}^{(\gamma_5,e)}[t] = V[t] \tilde{\Phi}^{(\gamma_5,e)}[t] V[t]^{\dagger},
\end{equation}
where $\tilde{\Phi}^{\gamma_5,e}[t]$ is the optimal elemental built from the optimal profile as given by Eq. \eqref{eqn:OptimalElementalMatrix} and $e=0,1$ denote the ground and first excited states. In the quark model this particular operator corresponds to a spin singlet and for this reason it is convenient to define the projected operator
\begin{equation}
\tilde{\Gamma}^{(\gamma_5,e)}_{S}[t] = Tr\left[ \gamma_5 \tilde{\Gamma}^{\gamma_5,e}[t] \right],
\label{eqn:SWave_Gamma}
\end{equation}
where $\gamma_5$ is chosen to reflect the fact that this is a spin singlet state and the trace is taken only over the Dirac indices, so that $\tilde{\Gamma}^{(\gamma_5,e)}_{S}[t]$ only has space and color indices. For a fixed time $t$ and a point-like source at an arbitrary 3D position $\vec{z}$ given by $\phi_{\vec{x}} = h_0 \delta_{\vec{x},\vec{z}} $ ($h_0 \in \mathbb{C}^3$, $||h_0||_2 = 1$) one can define the vector
\begin{equation}
\Phi^{(\gamma_5,e)}(\vec{x},t) = \sum_{\vec{y}} \tilde{\Gamma}^{(\gamma_5,e)}_{S}[t]_{\vec{x},\vec{y}} \phi_{\vec{y}}\, .
\end{equation}
The spatial distribution of interest is obtained by averaging its norm squared over all time slices
\begin{equation}
\Psi^{(\gamma_5,e)}(\vec{x}) = \frac{1}{N_t}\sum_{t=0}^{N_t -1} ||\Phi^{(\gamma_5,e)}(\vec{x},t)||_2^2,
\end{equation}
where the norm is taken in color space such that $\Psi^{(\gamma_5,e)}(\vec{x})$ is only a function of position that can be calculated for each gauge configuration and normalized so that $\sum_{\vec{x}}\Psi^{(\gamma_5,e)}(\vec{x}) = 1$. Since the spin singlet has $S=0$ then the angular momentum must satisfy $L=0$ within the quark model, giving it an S-wave classification and justifying the subscript "S" in Eq. \eqref{eqn:SWave_Gamma}. One would therefore expect the spatial distribution of the resulting vector to have an S-wave behavior. This is tested by setting the point-like source to be in $\vec{z} = (12a,12a,12a)$ and explicitly calculating $\Psi^{(\gamma_5,e)}(\vec{x})$. Fig. \ref{fig:SpatialProfiles_G5} shows $\Psi^{(\gamma_5,e)}(\vec{x})$ ($e=0,1$ for ground and first excited state) in the $xy$-plane at $z=12a$ averaged over 2 gauge configurations, which agrees quite well with the mentioned expectation in terms of spherical symmetry and number of nodes. 

\begin{figure}[H]
\centering
\includegraphics[width=0.9\linewidth]{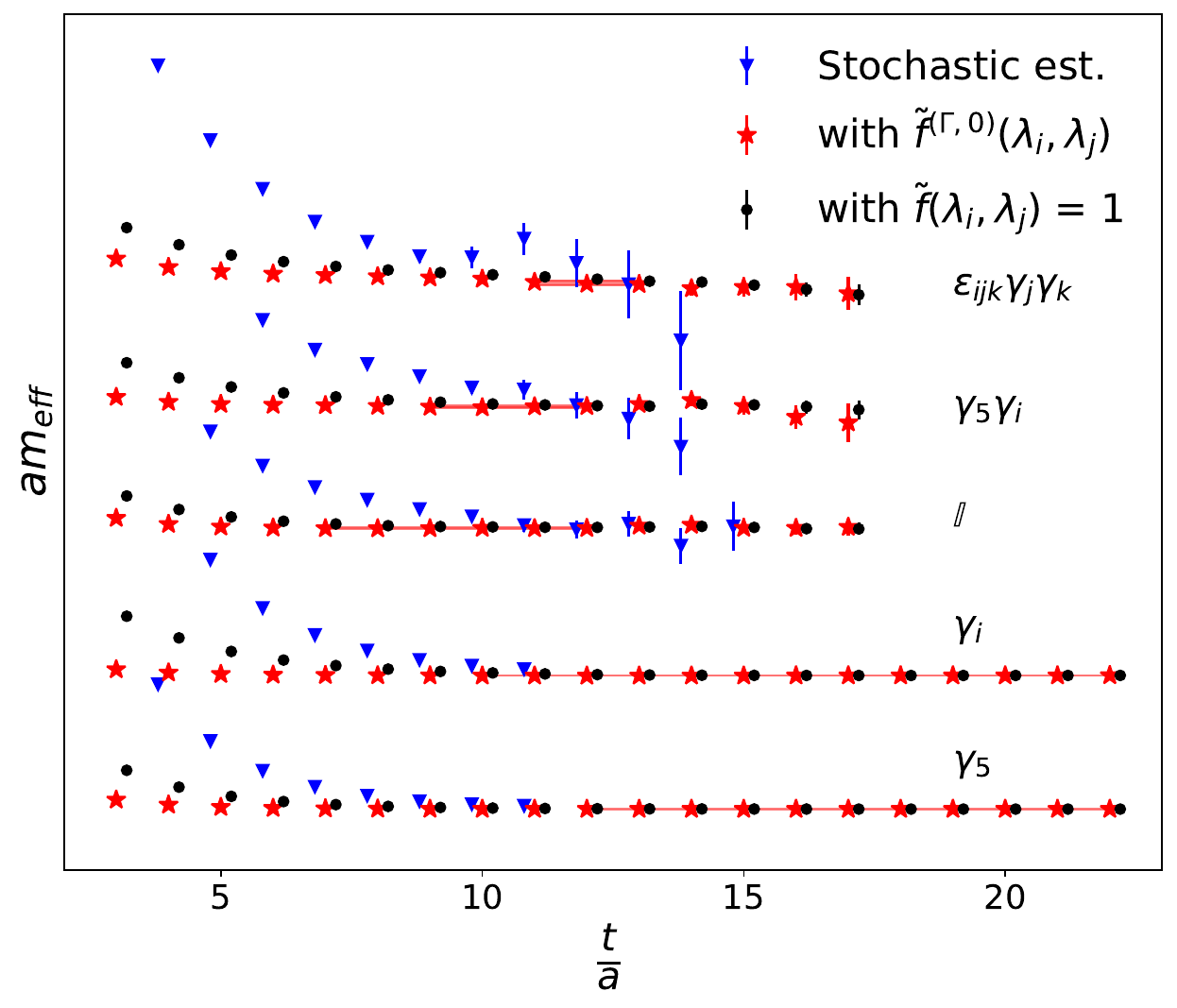}
\caption{Ground state effective masses of the local operators studied in this work using standard distillation, $\tilde{f}(\lambda_i, \lambda_j) = 1$, distillation with the optimal profiles $\tilde{f}^{(\Gamma,0)}(\lambda_i,\lambda_j)$ and stochastic trace estimation for each operator $\Gamma$. Plateaus of the ones using the optimal meson profiles are displayed. Different channels are displaced for clarity. Points of later times of the estochastic estimation results are omitted as to emphasize only their slower approach to the plateau.}
\label{fig:EffMass_Comparison}
\end{figure}

\noindent The extraction of effective masses for the different states, together with the corresponding optimal profiles, is first studied for the local operators listed in Table \ref{table:OperatorList}. Fig. \ref{fig:EffMass_Comparison} shows the ground state effective masses of some of these operators using three different methods; standard distillation, distillation with optimal meson profiles and stochastic trace estimation. The latter is done by evaluating the expression
\begin{align}
\pm \sum_{\vec{x},\vec{y}} \left\langle Tr\left[ \Gamma D^{-1}(x,y) \bar\Gamma D^{-1}(y,x) \right] \right\rangle_{U}  
\end{align}

\noindent using 16 $U(1)$ noise sources per configuration and a total of 2000 configurations. The source vectors are non-zero only on the source time-slice $y_0$, chosen randomly on each configuration. The $\Gamma=\gamma_5$ correlator requires one inversion per noise vector and configuration and for every additional $\Gamma$ one more such set of inversions is needed. This method of evaluating mesonic correlators is commonly used, e.g. recently in~\cite{Cali:2019enm}. The calculation is carried out using a variant of the program \verb+mesons+\footnote{Available at https://github.com/to-ko/mesons}. For all local operators standard distillation leads to a significant decrease of excited state contamination compared to the traditional stochastic estimation and the optimal profile further improves on this. For $\Gamma = \mathbb{I}, \gamma_5 \gamma_i, \epsilon_{ijk}\gamma_i \gamma_j$ the use of both variants of distillation also reduces the notorious presence of noise that the stochastic estimation displays after the excited state contamination seems to have been suppressed. Table \ref{table:PlateauAverages_Local} shows the corresponding plateau averages \footnote{Plateau averages are weighted by the inverse errors squared} for standard and optimized distillation. For all cases, use of distillation allows access to an effective mass plateau and in addition, the use of the optimal meson profiles notably decreases excited state contamination, leading to earlier plateaus. Notable examples are the $\gamma_5$ and $\gamma_i$ operators, where the length of the plateaus more than doubles. This improvement clearly speaks in favor of such profiles over standard distillation, especially since the number of necessary inversions remains unchanged.

\begin{table}[H]
\centering
\begin{tabular}{|c c c c|} 
 \hline
 $J^{PC}$ & $\Gamma$ & Plateau interval & $am$ \\ [0.5ex] 
 \hline\hline
 $0^{-+}$ & $\gamma_5$ & 12-22 & 0.74986(8)\\
          &            & 19-22 & 0.74986(8)\\
 $1^{--}$ & $\gamma_i$ & 10-22 & 0.8590(1)\\ 
          &            & 19-22 & 0.8590(1)\\
 $0^{++}$ & $\mathbb{I}$ & 7-12 & 1.0796(4)\\
          &              & 10-13 & 1.080(7)\\
 $1^{++}$ & $\gamma_5 \gamma_i$ & 9-12 & 1.1291(7)\\
          &                     & 11-13 & 1.130(1)\\
 $1^{+-}$ & $\epsilon_{ijk}\gamma_j \gamma_k$ & 11-13 & 1.130(2)\\ 
          &                                   & 16-18 & 1.123(7)\\[1ex]
 \hline
\end{tabular}
\caption{Ground state effective mass plateau averages for the local operators. The first row corresponds to the case with optimal meson distillation profiles while the second row corresponds to using standard distillation.}
\label{table:PlateauAverages_Local}
\end{table}

\begin{figure}[H]
\centering
\includegraphics[width=0.9\linewidth]{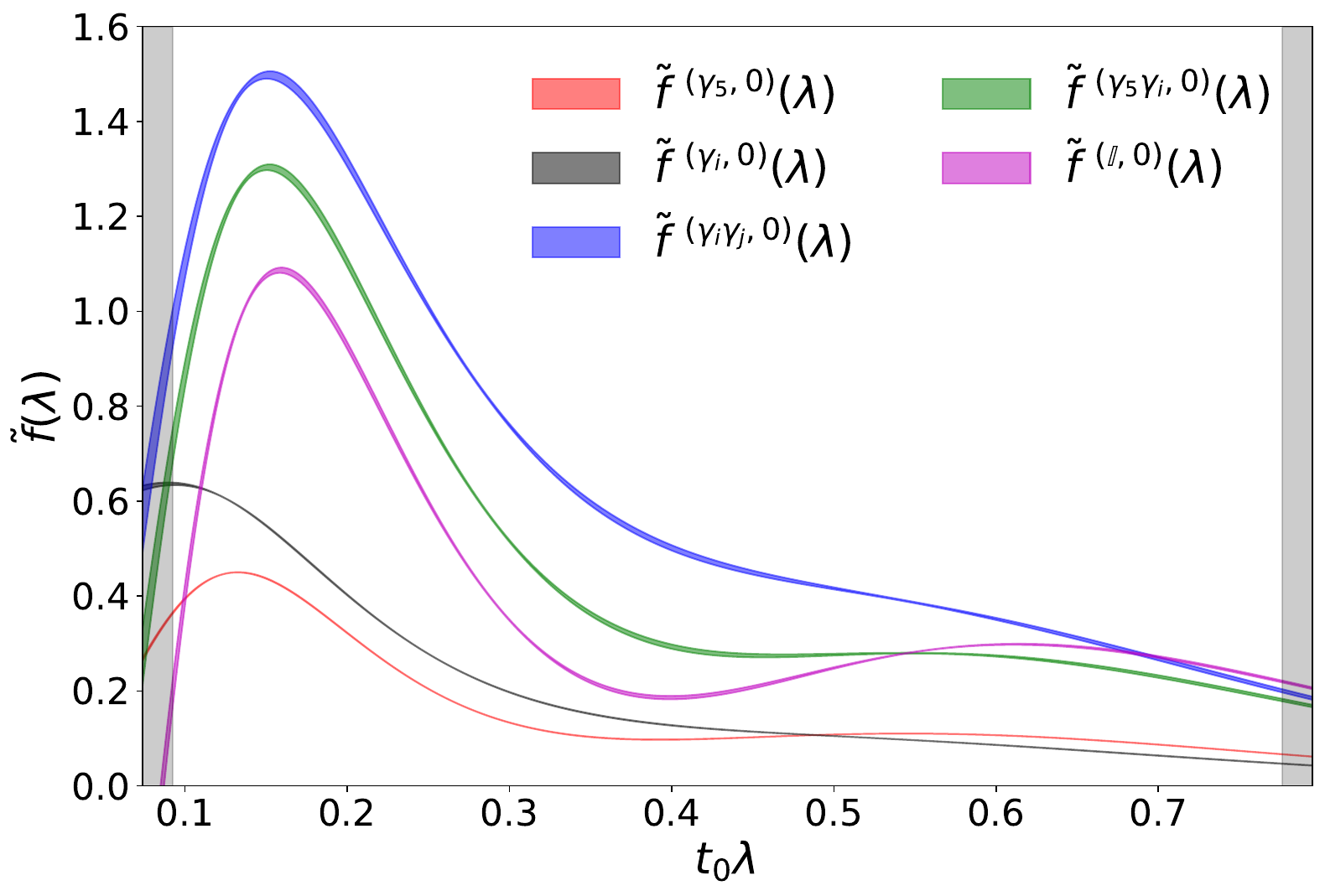}
\caption{Optimal meson distillation profile for the ground state of the local operators.}
\label{fig:LocalGround_Profiles}
\end{figure}

\noindent The resulting profiles for the ground states of the local operators are displayed in Fig. \ref{fig:LocalGround_Profiles}. All display the previously mentioned main characteristic: a modulated suppression of higher Laplacian eigenmodes. Although the heights of the different profiles differ, their overall shapes seem to be very similar, especially for the $\mathbb{I}$, $\gamma_5 \gamma_i$ and $\epsilon_{ijk}\gamma_j \gamma_k$ operators which suppress higher modes less than the $\gamma_5$ and $\gamma_i$ operators. The latter two have very similar profiles. In  all cases the profiles are different from the flat Heaviside function of standard distillation.  

\begin{figure}[H]
\centering
\includegraphics[width=0.9\linewidth]{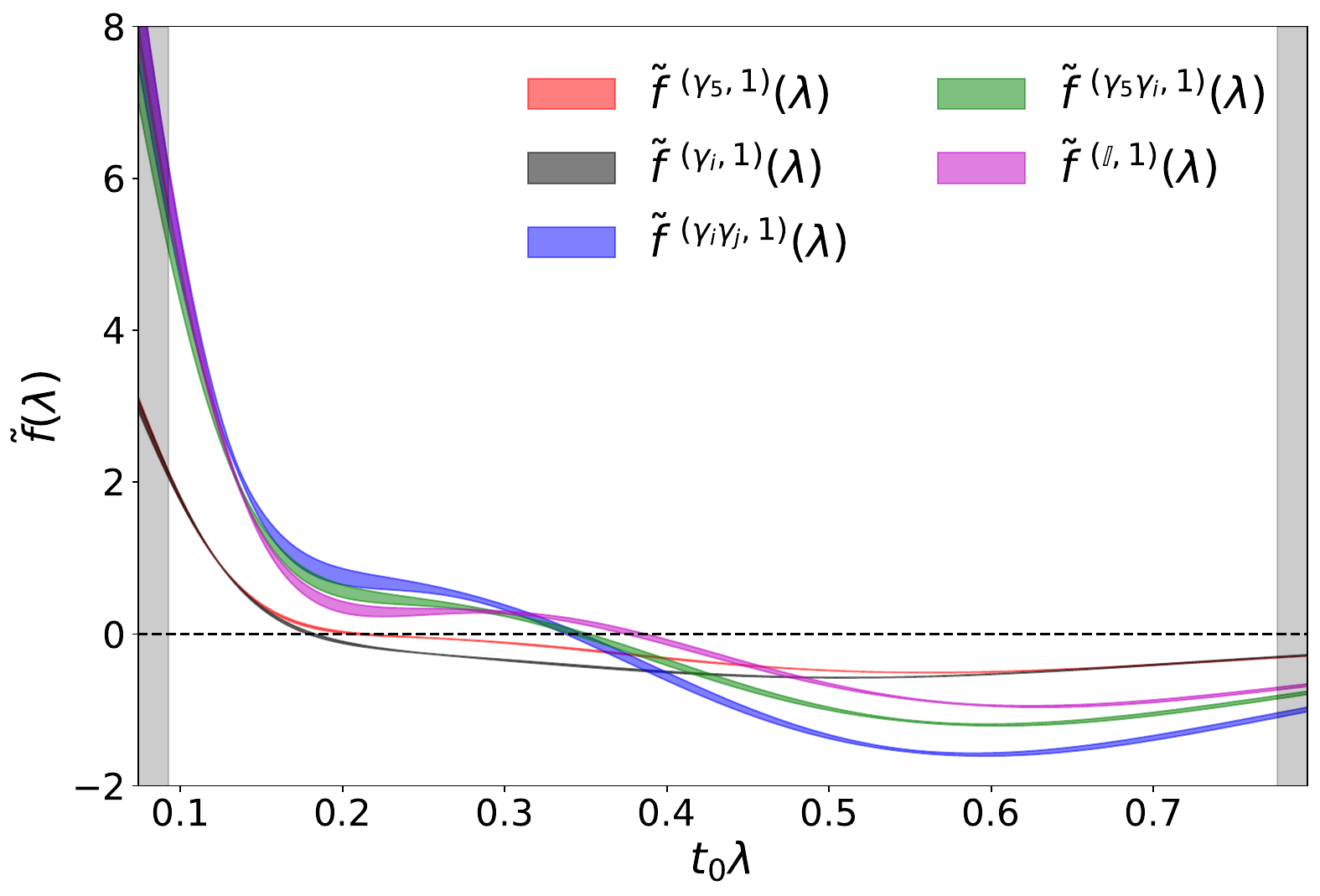}
\caption{Optimal meson distillation profile for the first excited state of the local operators.}
\label{fig:LocalFirst_Profiles}
\end{figure}

\noindent Fig. \ref{fig:LocalFirst_Profiles} shows the optimal meson profiles for the first excited states created by the local operators, where again increased dependence on higher eigenmodes is observed. As for the ground state, the $\mathbb{I}$, $\gamma_5 \gamma_i$ and $\epsilon_{ijk}\gamma_j \gamma_k$ are very similar in shape just as the $\gamma_5$ and $\gamma_i$ are to each other, the latter group more strongly suppressing the higher eigenmodes than the former and having an earlier node. For the case of the second excited state the same similarity between groups of operators is observed regarding shape, location of the two nodes and the level of suppression of eigenmodes.

\begin{figure}[H]
\centering
\includegraphics[width=0.9\linewidth]{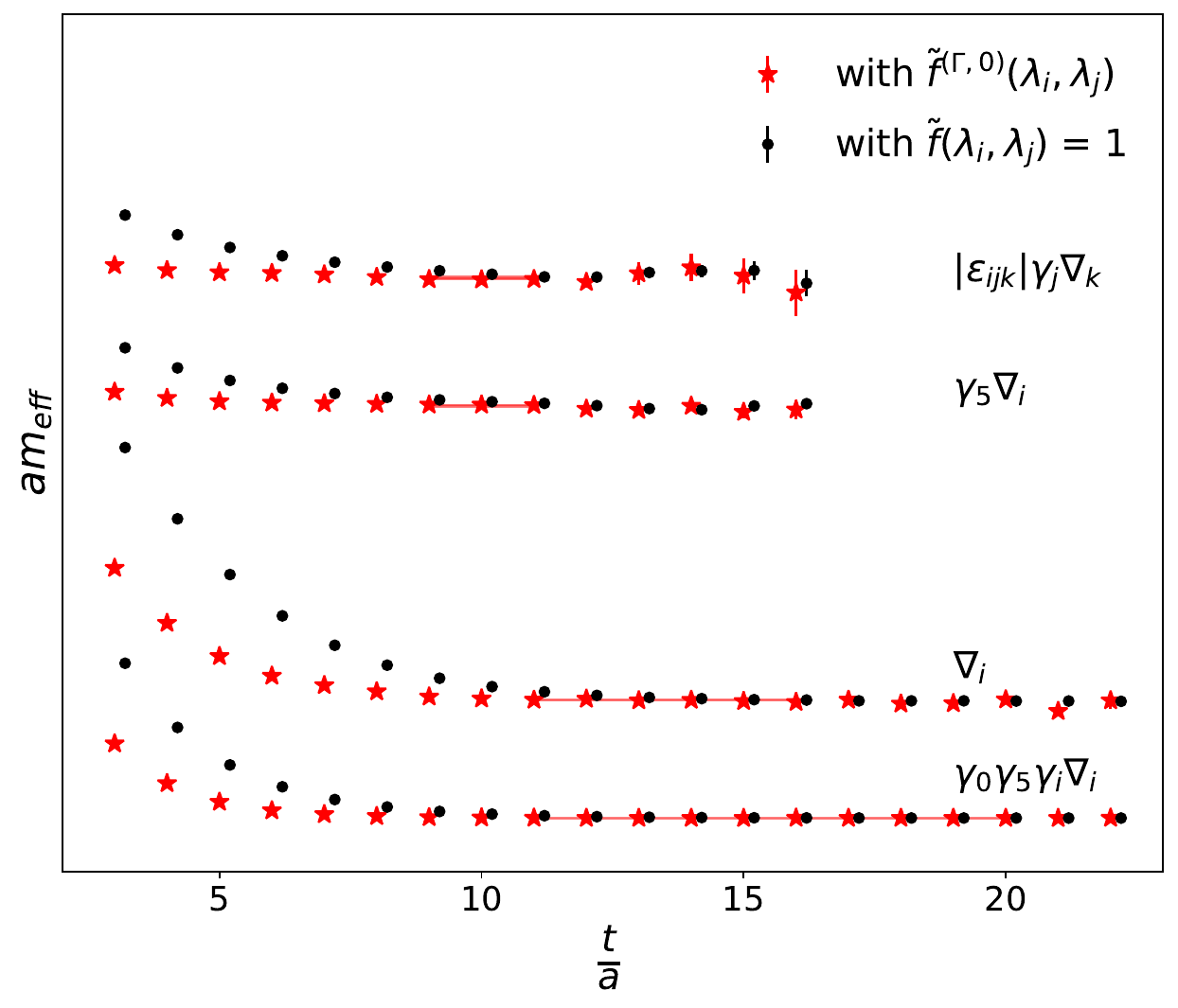}
\caption{Ground state effective masses of some of the derivative-based operators using both standard distillation, $\tilde{f}(\lambda_i, \lambda_j) = 1$, and the one with the optimal profiles $\tilde{f}^{(\Gamma,0)}(\lambda_i,\lambda_j)$ for each operator $\Gamma$. Plateaus of the optimal meson profiles are displayed. Different channels are displaced for clarity.}
\label{fig:EffMass_Comparison_Deriv}
\end{figure}

\begin{figure}[H]
\centering
\includegraphics[width=0.9\linewidth]{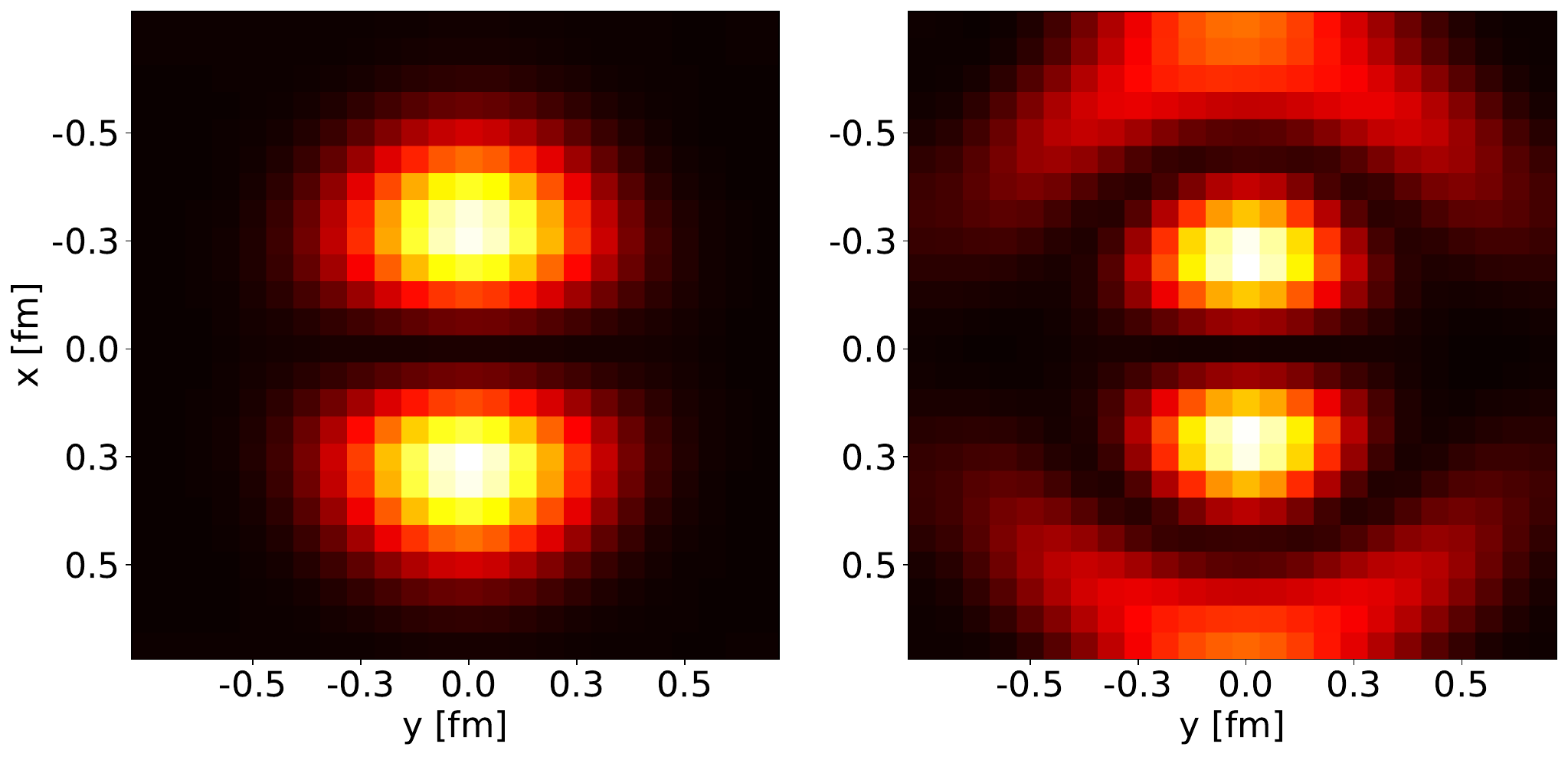}
\caption{Spatial distributions $\Psi^{(\gamma_5\nabla_1,0)}(\vec{x})$ and $\Psi^{(\gamma_5\nabla_1,1)}(\vec{x})$. The coloring changes from dark (zero) to light as the values of the profile increase.}
\label{fig:SpatialProfiles_G5D1}
\end{figure}

\noindent The effective masses and optimal meson distillation profiles are also extracted for the derivative-based operators. As with the local operators, these profiles yield a significant improvement over standard distillation. Fig. \ref{fig:EffMass_Comparison_Deriv} shows the ground state effective masses for some of the derivative-based operators studied. The gain from the optimal profiles can be seen as clearly as for the local operators. Table \ref{table:PlateauAverages_Deriv} shows the corresponding plateau averages. Notable improvement can be seen for example with the operator $\gamma_0 \gamma_5 \gamma_i \nabla_i$, where the length of the plateau more than doubles compared to standard distillation. This speaks for the importance of these profiles to make sure the relevant elementals use the eigenvectors in an optimal way. As was done for the $\gamma_5$ operator for the S-wave, the spatial distribution of the optimal $\tilde{\Gamma}[t]$ corresponding to the spin singlet P-wave, $J^{PC} = 1^{+-}$ can be visualized. It is studied via $\Gamma = \gamma_5 \nabla_i$ when one derivative is used. The same procedure used for the $\gamma_5$ operator is employed for $\gamma_5 \nabla_i$, yielding the spatial distributions shown in Fig. \ref{fig:SpatialProfiles_G5D1} for $i=1$.

\begin{table}[H]
\centering
\begin{tabular}{|c c c c|} 
 \hline
 $J^{PC}$ & $\Gamma$ & Plateau interval & $am$\\ [0.5ex] 
 \hline\hline
 $0^{-+}$ & $\gamma_0 \gamma_5 \gamma_i \nabla_i$ & 12-21 & 0.74989(8)\\
          &                                       & 20-22 & 0.74987(8)\\
 $1^{--}$ & $\nabla_i$ & 13-16 & 0.8589(3)\\ 
          &            & 17-20 & 0.8592(1)\\
 $1^{++}$ & $\epsilon_{ijk}\gamma_j \nabla_k$ & 9-13  & 1.1290(7)\\
          &                                   & 13-15 &  1.129(9)\\         
 $1^{+-}$ & $\gamma_5 \nabla_i$ & 9-11 & 1.1348(8)\\
          &                     & 13-15 & 1.134(2)\\
 $2^{++}$ & $|\epsilon_{ijk}| \gamma_j \nabla_k$ & 9-11 & 1.152(2)\\ 
          &                                      & 11-12 & 1.154(2)\\[1ex]
 \hline
\end{tabular}
\caption{Ground state effective mass plateau averages for some of the derivative-based operators used. The first row corresponds to the case with optimal meson distillation profiles while the second row corresponds to using standard distillation.}
\label{table:PlateauAverages_Deriv}
\end{table}

\begin{figure}[H]
\centering
\includegraphics[width=0.9\linewidth]{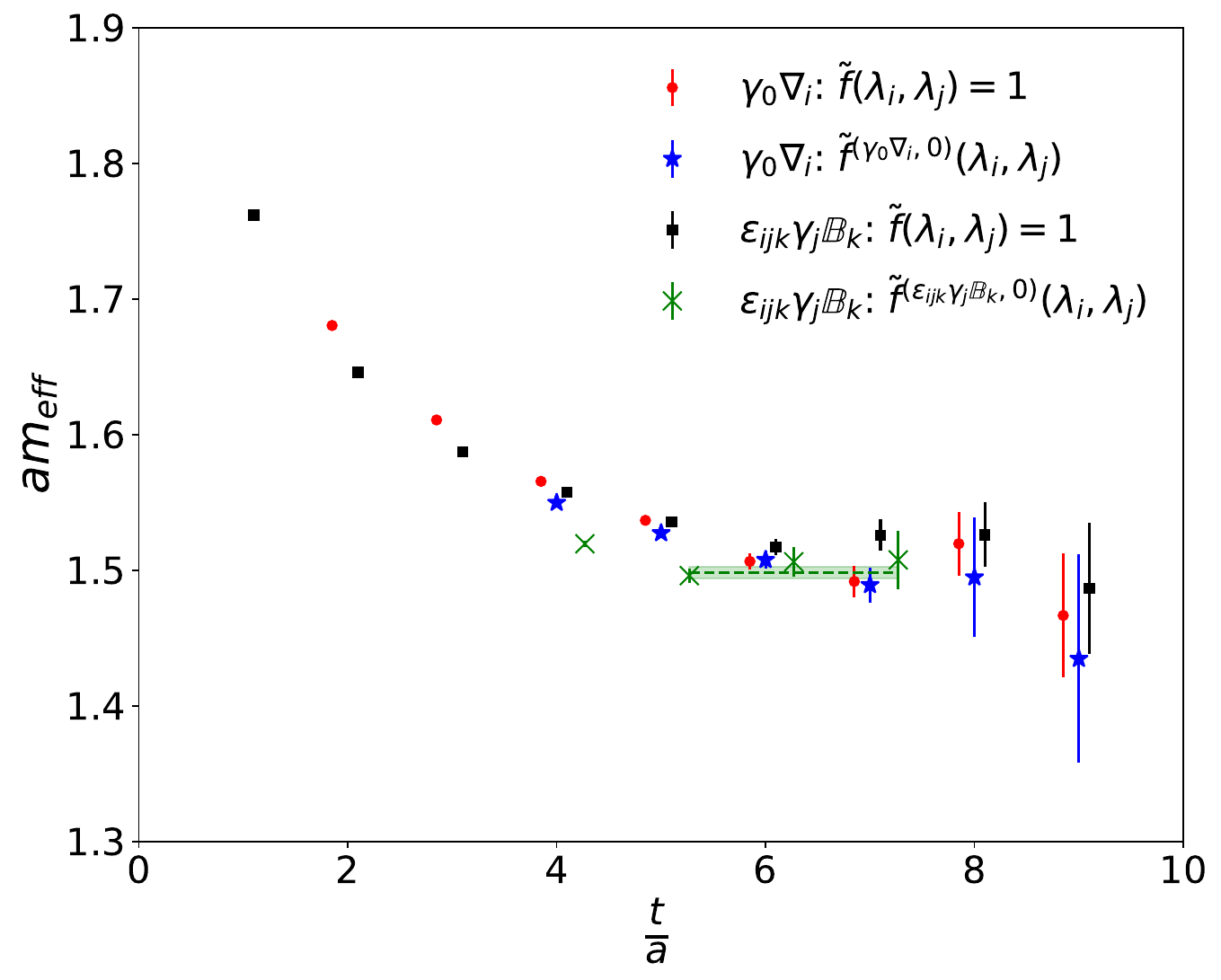}
\caption{Ground state effective masses of $1^{-+}$ operators using both standard distillation and the optimal meson profiles. The data points are displaced horizontally for clarity and omitted if the signal is lost to the noise.}
\label{fig:1mp_Comparison}
\end{figure}

\noindent Of special interest is the $1^{-+}$ channel, an exotic quantum number that can be modeled as a hybrid meson which includes explicit gluonic degrees of freedom combined with quark and anti-quark components \cite{Liu2012}. The operators $\gamma_0 \nabla_i$ and $\epsilon_{ijk}\gamma_j \mathbb{B}_k$ are used to study this channel, where the latter operator includes the field-strength tensor via $\mathbb{B}_k$ for gluonic excitation.  Fig. \ref{fig:1mp_Comparison} shows the ground state effective mass for both operators with and without the optimal meson distillation profiles. Results corresponding to the optimal profiles are plotted starting at $t_G + a$ and shown only when the error is smaller than the signal. As expected, the optimal profile leads to a decrease of excited state contamination, more substantial for the $\epsilon_{ijk}\gamma_j \mathbb{B}_k$ operator than for the $\gamma_0 \nabla_i$ one. Second, when considering the optimal profile for both operators there is a clear difference in behavior of the effective mass which points to $\epsilon_{ijk}\gamma_j \mathbb{B}_k$ having most overlap with the lowest energy eigenstate. Since $\epsilon_{ijk}\gamma_j \mathbb{B}_k$ has an explicit gluonic excitation, unlike $\gamma_0 \nabla_i$, this operator would intuitively have better overlap with the energy eigenstate if the state is indeed a hybrid. This scenario is favored in other studies of this channel \cite{Dudek2010, Liu2012, Cheung2016, MankeOperators, Dudek2009}.

\begin{figure}[H]
\centering
\includegraphics[width=0.9\linewidth]{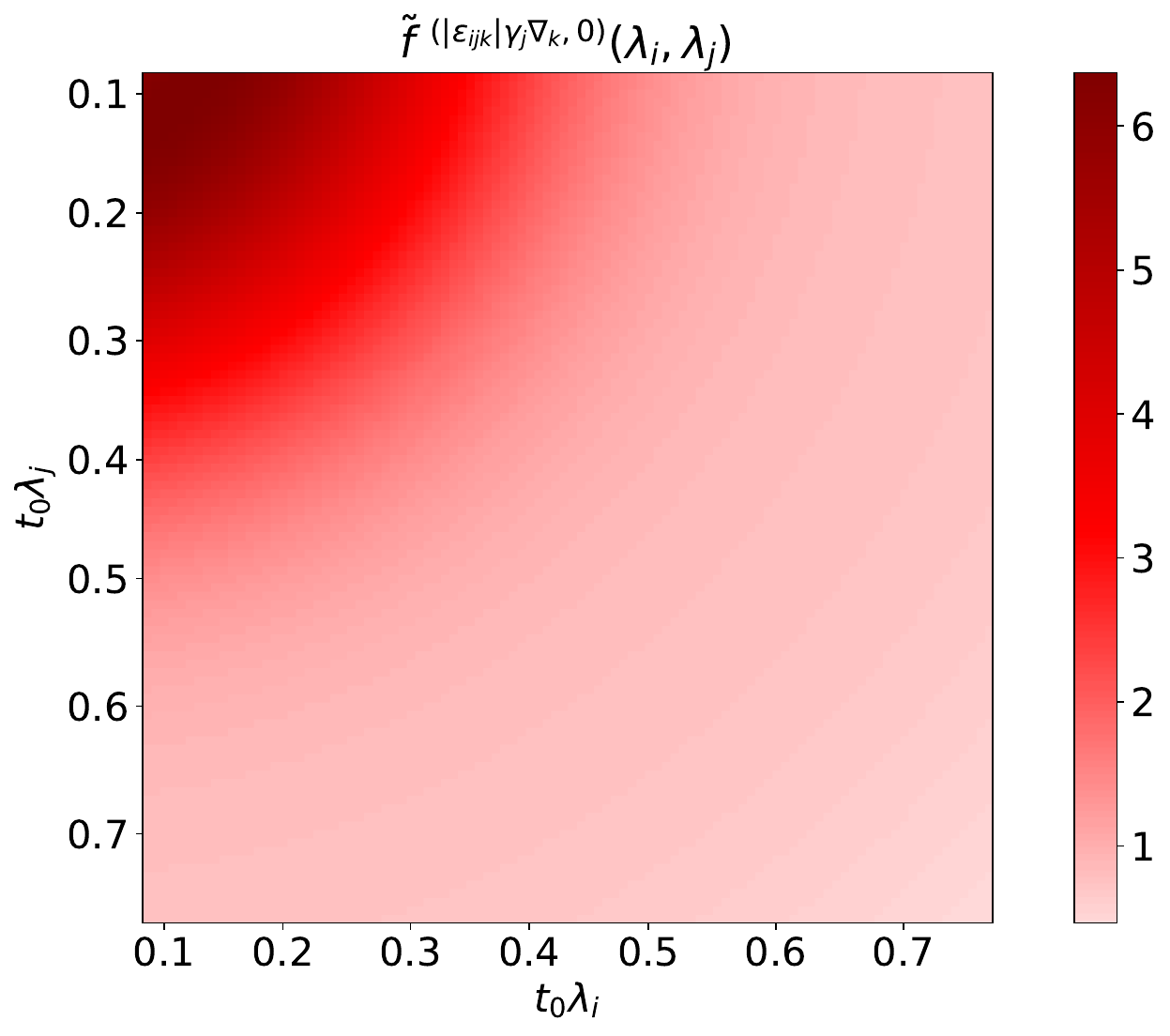}
\caption{Optimal meson distillation profile for the ground state of the $|\epsilon_{ijk}|\gamma_j \nabla_k$ operator.}
\label{fig:SijkGjDk_Ground_Profile}
\end{figure}

\begin{figure}[H]
\centering
\includegraphics[width=0.9\linewidth]{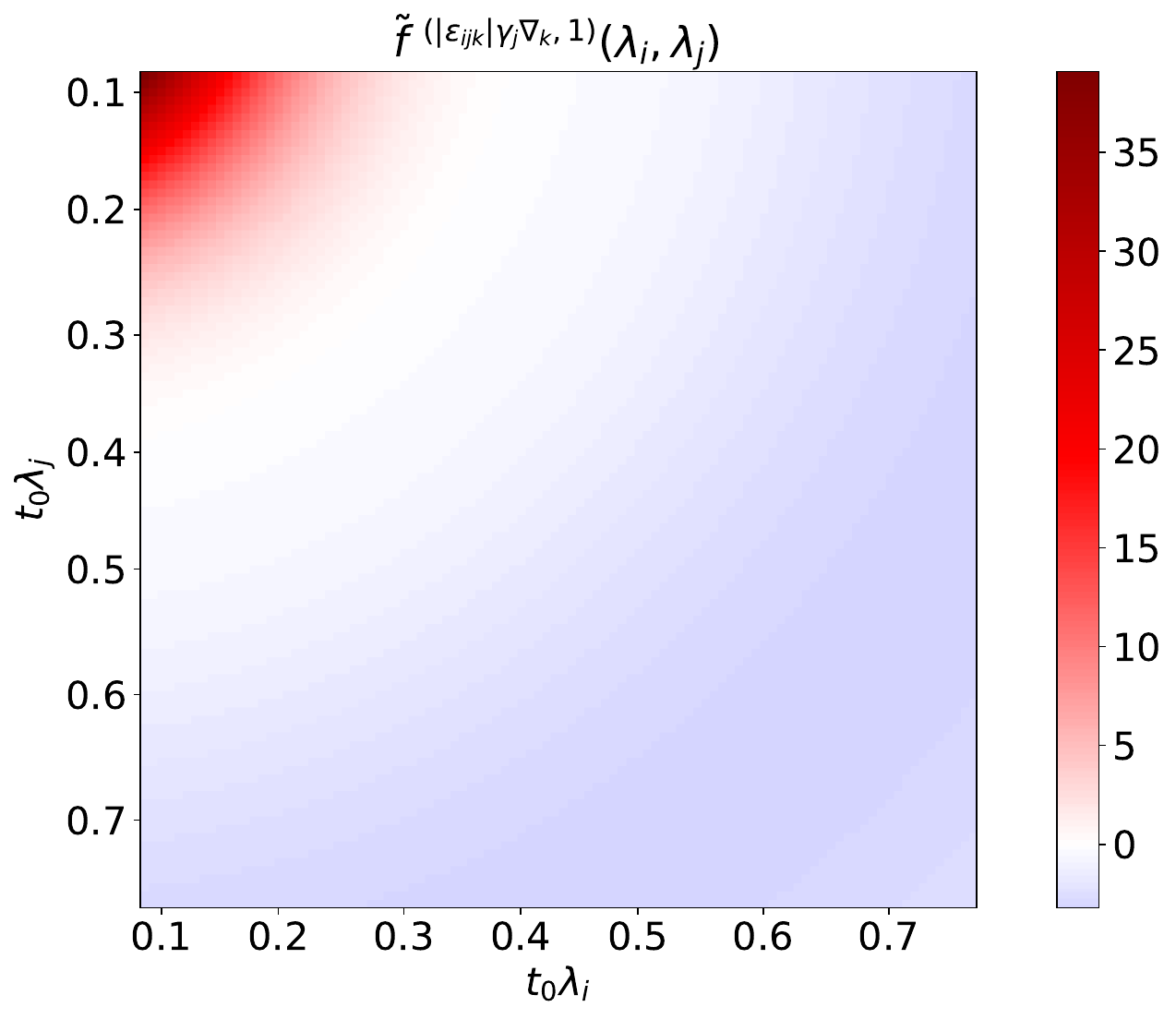}
\caption{Optimal meson distillation profile for the first excited state of the $|\epsilon_{ijk}|\gamma_j \nabla_k$ operator.}
\label{fig:SijkGjDk_First_Profile}
\end{figure}

\begin{figure}[H]
\centering
\includegraphics[width=0.96\linewidth]{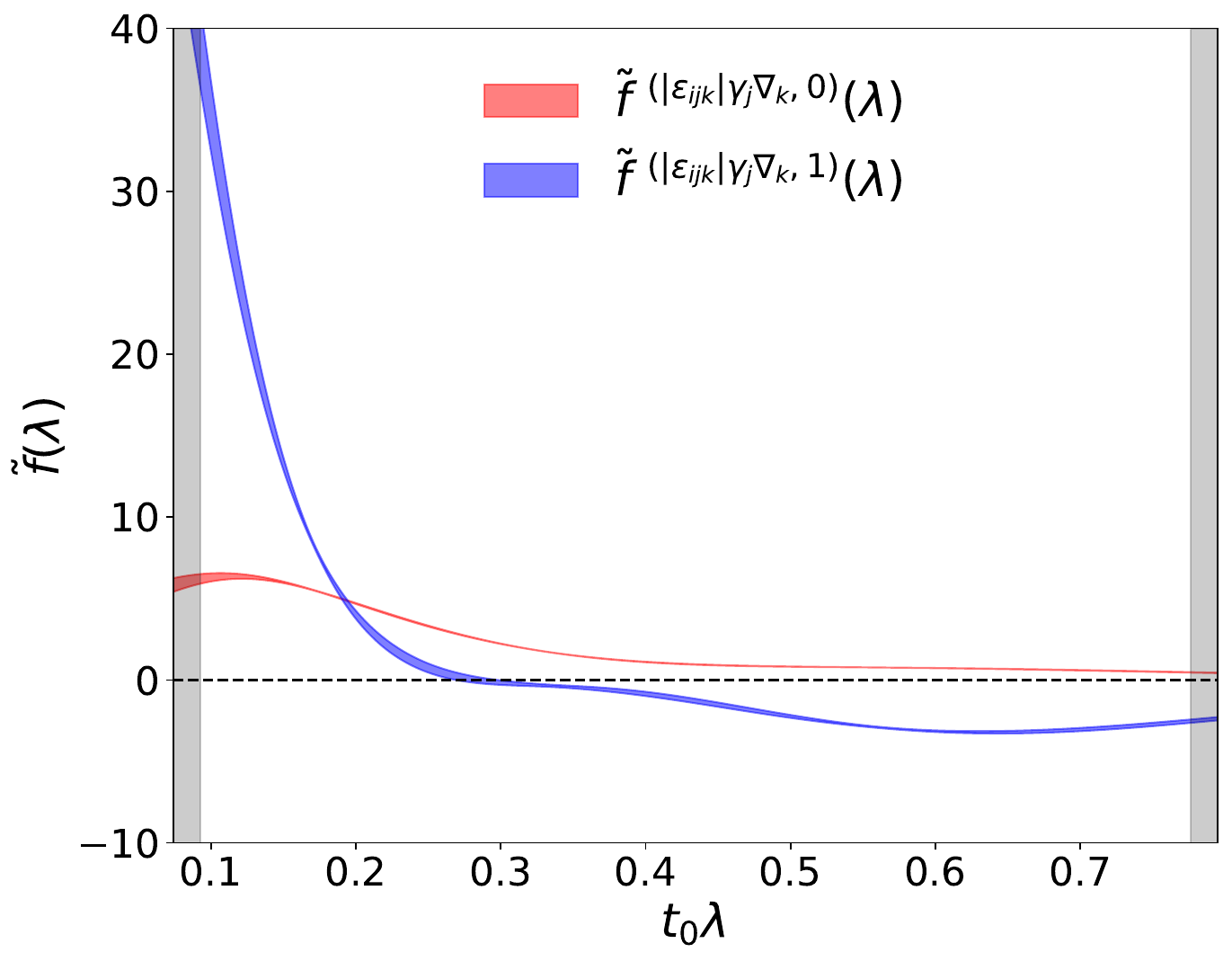}
\caption{Optimal meson distillation profiles for the ground and first excited state of the $|\epsilon_{ijk}|\gamma_j \nabla_k$ operator when $\lambda_i = \lambda_j$.}
\label{fig:SijkGjDk_Ground_Profile_Diag}
\end{figure}

\noindent Optimal meson profiles are extracted for all derivative-based operators used. The first optimal profile shown in Fig. \ref{fig:SijkGjDk_Ground_Profile} corresponds to the ground state created by the operator $|\epsilon_{ijk}|\gamma_j \nabla_k$. 
The main contribution to the profile comes clearly from pairs of low eigenvalues, as expected in distillation, and exhibits an approximately radially symmetric decay with increasing eigenvalues. As with the local operators, no nodes are observed for the ground state. Fig. \ref{fig:SijkGjDk_First_Profile} shows the optimal profile for the first excited state, where a single node can be seen and, while higher values of eigenvalues are less suppressed than in the ground state profile, the overall tendency of lower eigenmodes contributing more than higher ones still remains. To compare these profiles with those obtained for local operators one can take the one-dimensional profiles $\tilde{f}^{(|\epsilon_{ijk}|\gamma_j \nabla_k,0)}(\lambda_i, \lambda_i)$ and $\tilde{f}^{(|\epsilon_{ijk}|\gamma_j \nabla_k,1)}(\lambda_i, \lambda_i)$, shown in Fig. \ref{fig:SijkGjDk_Ground_Profile_Diag} with error bands. The notation $\tilde{f}^{(|\epsilon_{ijk}|\gamma_j \nabla_k,0)}(\lambda)$ should be understood as $\tilde{f}^{(|\epsilon_{ijk}|\gamma_j \nabla_k,0)}(\lambda, \lambda)$ and the same for the first excited state. The suppression of higher eigenmodes, together with the node for the first excited state, are clearly visible with notable resemblance to the local operators. 

\newpage

\begin{figure}[H]
\centering
\includegraphics[width=0.95\linewidth]{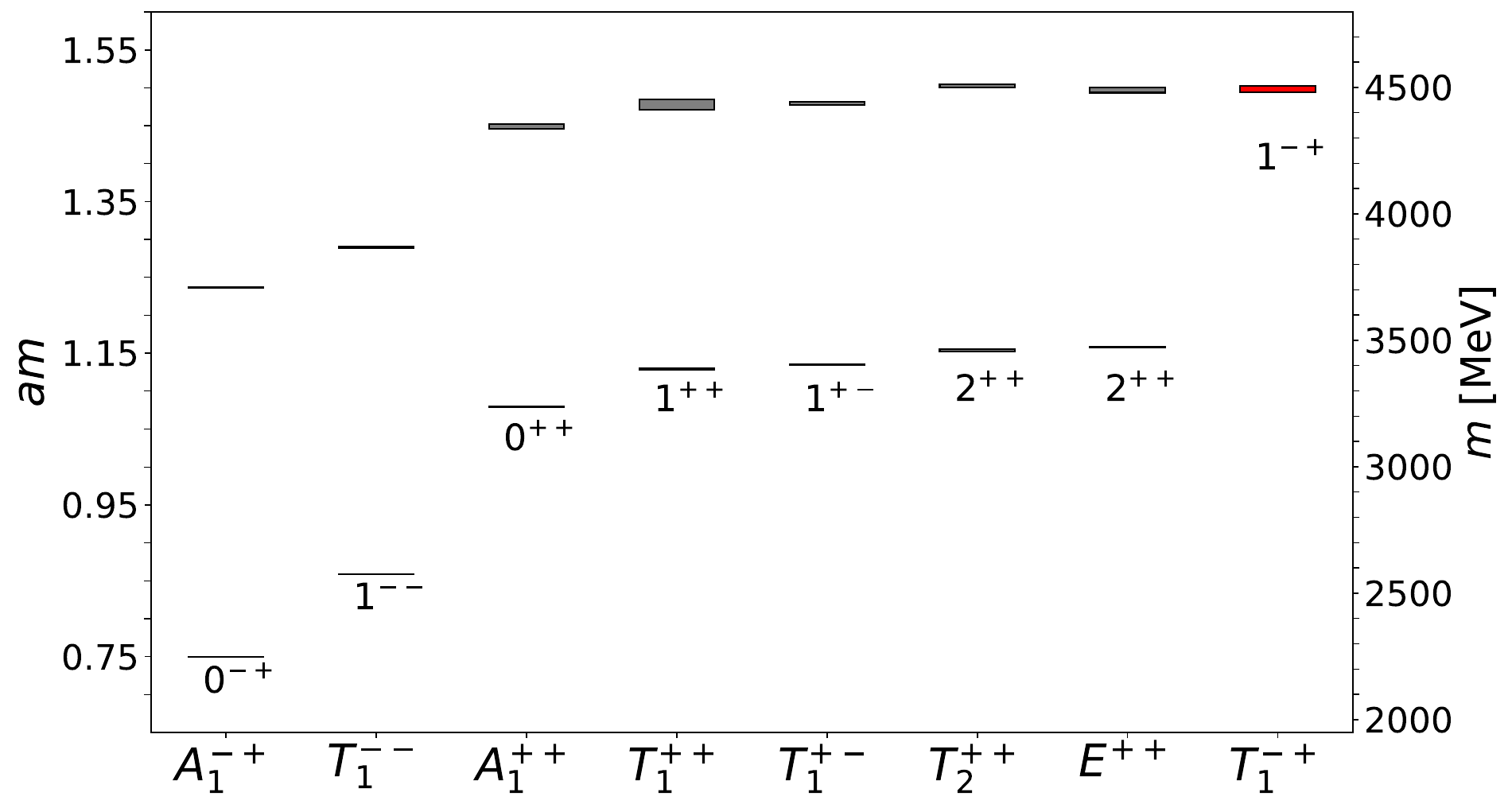}
\caption{Charmonium spectrum for all the studied channels.}
\label{fig:Charmonium_Spectrum}
\end{figure}

\noindent At this point it is possible to gather the results obtained with both local and derivative-based operators to get a clear picture of the charmonium spectrum accessible through the use of the optimal meson distillation profiles when considering only the connected correlations. Table \ref{table:PlateauAverages_All} displays the plateau averages of the ground and first excited state effective masses for all the operators studied in this work, complementing the contents of tables \ref{table:PlateauAverages_Local} and \ref{table:PlateauAverages_Deriv}. The $1^{-+}$ channel only includes the ground state since it was the only clearly accessible one. The remarkable agreement between different operators of the same $J^{PC}$ serves as a non-trivial test. The slight tension between the $E^{++}$ and $T_2^{++}$ ground states can be explained via lattice artifacts since these two should coincide when taking the continuum limit. Fig. \ref{fig:Charmonium_Spectrum} displays the ground and first excited state for each fixed irrep, where the values come from the operators that displayed the best signal, and the spin assignment for the ground state is given by the label under each data point. The access to a first excited state purely by the inclusion of  meson distillation profiles further demonstrates their usefulness. Note the heirarchy of states computed matches the pattern seen in nature, where there are eight narrow charmonium resonances below the $D\bar{D}$ threshold. In this investigation, which has no light dynamical quarks, this threshold is
absent. The ground state for the $1^{-+}$ channel is shown in red to emphasize its spin-exotic nature.

\newpage

\begin{table}[H]
\centering
\begin{tabular}{|c c c c|} 
 \hline
 $J^{PC}$ & $\Gamma$ & Plateau interval & $am$\\ [0.5ex] 
 \hline\hline
 $0^{-+}$ & $\gamma_5$ & 12-22 & 0.74986(8)\\
          &            & 5-9   & 1.2363(6)\\
          & $\gamma_0 \gamma_5 \gamma_i \nabla_i$ & 12-21 & 0.74989(8)\\
          &            & 6-9   & 1.238(2)\\
          & $\gamma_i \mathbb{B}_i$ & 10-22 & 0.74991(9)\\
          &                         & 7-8 & 1.21(2)\\
 $1^{--}$ & $\gamma_i$ & 10-22 & 0.8590(1)\\ 
          &            & 5-8   & 1.2898(7)\\
          & $\nabla_i$ & 13-16 & 0.8589(3)\\
          &            & 6-8   & 1.325(4)\\
          & $\gamma_5 \mathbb{B}_i$ & 11-17 & 0.8587(6)\\
          &                         & 5-6 & 1.43(1)\\
 $0^{++}$ & $\mathbb{I}$ & 7-12 & 1.0794(4)\\
          &              & 6-7  & 1.443(3)\\
          & $\gamma_i \nabla_i$ & 7-13 & 1.0792(3)\\
          &                     & 6-7 & 1.448(3)\\
 $1^{++}$ & $\gamma_5 \gamma_i$ & 9-12 & 1.1287(9)\\
          &                     & 7-9  & 1.477(6)\\
          & $\epsilon_{ijk} \gamma_j \nabla_k$ & 9-13 & 1.1290(7)\\
          &                                    & 5-7  & 1.479(2)\\
 $1^{+-}$ & $\epsilon_{ijk}\gamma_j \gamma_k$ & 11-13 & 1.130(2)\\ 
          &                                   & 6-8   & 1.506(3)\\
          & $\gamma_5 \nabla_i$ & 9-11 & 1.1348(8)\\
          &                     & 5-7  & 1.479(2)\\
 $2^{++}$ & $|\epsilon_{ijk}| \gamma_j \nabla_k$ & 9-11 & 1.153(1)\\
          &                                      & 5-7 & 1.493(2)\\
          & $\mathbb{Q}_{ijk} \gamma_j \nabla_k$  & 8-10 & 1.1560(8)\\
          &                                       & 5-7 & 1.496(2)\\
 $1^{-+}$ & $\epsilon_{ijk}\gamma_j \mathbb{B}_k$ & 5-7 & 1.498(4)\\[1ex]
 \hline
\end{tabular}
\caption{Mass plateau averages for all operators with optimal distillation profiles $\tilde{f}^{(\Gamma,e)}(\lambda_i,\lambda_j)$ used in this work. The first row of each operator corresponds to the ground state ($e=0$) while the second line corresponds to the first excited state ($e=1$).}
\label{table:PlateauAverages_All}
\end{table}

\begin{table}[H]
\centering
\begin{tabular}{|c c c|} 
 \hline
 $\Delta M$ & Ground state & 1\textsuperscript{st} Excited state \\ [0.5ex] 
 \hline\hline
 $\Delta m_{HF}$ & $0.10916(7)$ & $0.0535(3)$ \\
 $\Delta m_{1P-1S}$ & $0.3066(6)$ & $0.208(3)$\\
 $\Delta m_{S-O}$ & $0.0261(5)$& $0.018(3)$\\
 $\Delta m_{tensor}$ & $0.0079(3)$& $0.005(2)$\\
 $\Delta m_{1PHF}$ & $0.0035(8)$ & $0.005(3)$\\[1ex]
 \hline \hline 
 $\Delta M$ & Ground state [MeV]& 1\textsuperscript{st} Excited state [MeV]\\
 \hline \hline
 $\Delta m_{HF}$ & 327.3 $\pm$ 0.2 $\pm$ 5 & 160.5 $\pm$ 0.9 $\pm$ 2.6\\
 \hline
\end{tabular}
\caption{Different mass splittings in lattice units studied in this work for the ground and first excited states. The last row corresponds to the hyperfine splitting in physical units where the first error ignores the error of the lattice spacing while the second one takes it into account.}
\label{table:MassSplittings}
\end{table}

\newpage

\noindent A set of quantities of interest are the mass splittings. Their definitions are taken from \cite{DeTar2019} and are 
\begin{align}
\Delta m_{HF} &= m_{J/\Psi} - m_{\eta_c}\\
\Delta m_{1P-1S} &= m_{\overline{1P}} - m_{\overline{1S}}\\
m_{\overline{1P}} &= \frac{1}{9}\left( m_{\chi_{c0}} + 3m_{\chi_{c1}} + 5m_{\chi_{c2}} \right)\\
m_{\overline{1S}} &= \frac{1}{4} \left(  m_{\eta_c} + 3m_{J/\Psi}\right)\\
\Delta m_{S-O} &= \frac{1}{9}\left( 5m_{\chi_{c2}} - 3m_{\chi_{c1}} - 2m_{\chi_{c0}} \right)\\
\Delta m_{tensor} &= \frac{1}{9} \left( 3m_{\chi_{c1}} - m_{\chi_{c2}} - 2m_{\chi_{c0}} \right)\\
\Delta m_{1PHF} &=  m_{\overline{1P}} - m_{h_c} ,
\end{align}
where $\Delta m_{HF}$ is the $1S$ hyperfine splitting, $\Delta m_{1P-1S}$ is the spin-average $1P-1S$ splitting, $\Delta m_{S-O}$ is the spin-orbit splitting, $\Delta m_{tensor}$ is the tensor splitting and $\Delta m_{1PHF}$ is the P-wave hyperfine splitting. They are calculated using the assignment shown in Table \ref{table:OperatorList} and from the operator that yields the mass with smallest error in lattice units. In all cases only iso-vector masses are considered. Table \ref{table:MassSplittings} shows the mass splittings obtained for the ground and first excited states. Good precision is achieved in both cases thanks to the use of the optimal distillation profiles. The last row in Table \ref{table:MassSplittings} gives the hyperfine splitting in physical units along with two uncertainties. The first indicates solely the statistical precision while the second includes the systematic uncertainty arising from scale setting, which relates the lattice spacing to physical units. Recall the quark mass is about one half of the physical charm-quark mass so this comparison is made to emphasise the difference between the precision obtained by this lattice calculation and the overall scale setting. The scale uncertainty is significantly higher in both cases.

\begin{table}[H]
\centering
\begin{tabular}{|c c c c|} 
 \hline
 $J^{PC}$ & $\Gamma$ & $\tilde{f}^{(\Gamma,0)}(\lambda_i, \lambda_j) = 1$ & Optimal $\tilde{f}^{(\Gamma,0)}(\lambda_i, \lambda_j)$\\ [0.5ex] 
 \hline\hline
 $0^{-+}$ & $\gamma_5$ & $0.9272(3)$ & $0.9858(2)$ \\
 		  &            &             & $0.980(1)$  \\
          & $\gamma_0 \gamma_5 \gamma_i \nabla_i$ & $0.7035(4)$ & $0.8776(3)$\\
          & $\gamma_i \mathbb{B}_i$ & $0.8228(4)$& $0.8473(3)$\\
 $1^{--}$ & $\gamma_i$ & $0.8743(10)$& $0.9900(5)$\\
          & $\nabla_i$ & $0.4758(7)$& $0.742(2)$\\
          & $\gamma_5 \mathbb{B}_i$ & $0.556(1)$& $0.578(4)$\\
 $0^{++}$ & $\mathbb{I}$ & $0.944(4)$& $0.986(1)$\\
          & $\gamma_i \nabla_i$ & $0.946(5)$& $0.978(1)$\\
 $1^{++}$ & $\gamma_5 \gamma_i$ & $0.907(8)$& $0.982(5)$\\
          & $\epsilon_{ijk} \gamma_j \nabla_k$ & $0.86(1)$& $0.972(3)$\\
 $1^{+-}$ & $\epsilon_{ijk} \gamma_j \gamma_k$ & $0.77(7)$& $0.93(1)$\\
          & $\gamma_5 \nabla_i$ & $0.84(1)$& $0.970(5)$\\
 $2^{++}\left( T_2 \right)$ & $|\epsilon_{ijk}| \gamma_j \nabla_k$ & $0.850(8)$& $0.969(5)$\\
 $2^{++}\left( E \right)$   & $\mathbb{Q}_{ijk} \gamma_j \nabla_k$ & $0.858(8)$ & $0.981(3)$\\
 $1^{-+}$ & $\epsilon_{ijk}\gamma_j \mathbb{B}_k$ & $0.81(1)$ & $0.952(9)$\\[1ex]
 \hline
\end{tabular}
\caption{Fractional overlaps with the ground state using standard distillation ($\tilde{f}^{(\Gamma,0)}(\lambda_i, \lambda_j) = 1$ ) and the optimal meson profiles (Optimal $\tilde{f}^{(\Gamma,0)}(\lambda_i, \lambda_j)$). For the $\Gamma = \gamma_5$ bilinear, data for the first-excited state is included}
\label{table:OverlapTable}
\end{table}

\noindent Finally, it is important to numerically quantify how much the use of the optimal distillation profiles helps to create a state closer to the desired energy eigenstate for the different channels. To do this one first looks at the correlation function calculated for a given $\Gamma$ using standard distillation. It has a large $t$ limit given by
\begin{equation}
C(t) = 2c_{0} e^{-m_0\frac{T}{2}} \cosh \left( \left( \frac{T}{2} - t \right) m_0 \right),
\label{eqn:CorrCosh}
\end{equation}
where $m_0$ is the ground state mass for the chosen channel and the coefficient $c_0$ is the amplitude squared of the overlap between the state created by the meson operator involving $\Gamma$ and the lowest energy eigenstate in the channel. However, to compare it with the obtained data one must take into account the presence of excited state contamination. The correlation function is then given by
\begin{align}
C(t)= 2c_{0} e^{-m_0\frac{T}{2}}\cosh \left( \left( \frac{T}{2} - t \right) m_0 \right) + B_1(t)
\label{eqn:ContaminedNormalizedCosh}
\end{align}
where $B_1(t)$ contains the excited state contamination. One can define a normalized correlator $C^{\prime}(t)$ that satisfies $C^{\prime}(t_G) = 1$ as
\begin{align}
\begin{split}
C^{\prime}(t) &= \frac{C(t)}{C(t_G)}\\
&= \left(\frac{1 + B_2(t)}{1 + B_2(t_G)}\right) \frac{\cosh \left( \left( \frac{T}{2} - t \right) m_0 \right)}{\cosh \left( \left( \frac{T}{2} - t_G \right) m_0 \right)}
\end{split}
\label{eqn:NormalizedCosh}
\end{align}
with $B_2(t)$ defined as
\begin{align}
B_2(t) = \frac{B_1(t)e^{m_0 \frac{T}{2}}}{2c_0 \cosh \left( \left( \frac{T}{2} - t \right) m_0 \right)}.
\end{align}
From the spectral decomposition it is known that $B_1(t) < B_1(t_G)$ if $t_G < t < \frac{T}{2}$, which is the regime which is analyzed, so it holds that $A(t) = \frac{1 + B_2(t)}{1 + B_2(t_G)} < 1$ in this regime. In the mass plateau interval where $B_1(t) = 0$ one obtains $A_G = \frac{1}{1 + B_2(t_G)} < 1$ and this parameter quantifies the presence of excited state contamination at early times. If there is none then $B_1(t_G) = 0$ and $A_G = 1$, however $B(t_G) > 0$ means $0 < A_G < 1$. The closer $A_G$ is to $1$ the larger the suppression of excited state contamination, which is the goal of the optimized operators used in this work. To get the value of $A_G$ one can fit the normalized correlation data to the form of Eqn. \eqref{eqn:NormalizedCosh} in the plateau region where $B_1(t)$ is considered to be 0. This way only one fit parameter must be found. To simplify the analysis one can solve for an effective parameter $A_{eff}(t)$ as
\begin{align}
A_{eff}(t) = \frac{C(t)}{C(t_G)}\frac{\cosh \left( \left( \frac{T}{2} - t_G \right) m_0 \right)}{\cosh \left( \left( \frac{T}{2} - t \right) m_0 \right)}
\end{align}
and estimate $A_G$ as the average of $A_{eff}(t)$ in the previously mentioned plateau region. A similar procedure can be done for the generalized eigenvalues from the GEVP defined in Eqn. \eqref{eqn:Eigenvalues}. They have a large $t$ limit given by \cite{Irges2007}
\begin{equation}
\rho_{e}(t,t_G) = 2c_{e} e^{-m_e\frac{T}{2}} \cosh \left( \left( \frac{T}{2} - t \right) m_e \right),
\end{equation}
which is an identical form to the normalized correlation in the case of standard distillation. One can again account for the presence of excited state contamination and perform a fit of the data to the form of Eqn. \eqref{eqn:NormalizedCosh} by taking $e=0$ and replacing $\frac{C(t)}{C(t_G)}$ with $\frac{\rho_{e}(t,t_G)}{\rho_{e}(t_G,t_G)}$, which is equal to $\rho_{e}(t,t_G)$ due to normalization. By doing so via an average of $A_{eff}(t)$ in the region of the effective mass plateau where $B_1(t)$ is $0$ one gets the value that quantifies the suppression of excited state contamination just as for the case of standard distillation. Once the value of $A_G$ is known for the ground state of every operator analyzed in this work via both standard distillation and the one with the optimal profile it is possible to compare them as to quantify the improvement that the optimal profile brings. These values are displayed in Table \ref{table:OverlapTable}, where $A_G$ is denoted as the fractional overlap.
For the $\gamma_5$ bilinear, the fractional overlap from fitting the first-excited state is included. Again, a value close to unity indicates the profiles also accurately represent excited-state wavefunctions. Note no comparison with the standard distillation method is available, since only a single correlation function is generated from the $\gamma_5$ bilinear. As expected this fractional overlap is larger when using the optimal profiles compared to standard distillation. Remarkable improvement is obtained for a majority of the operators studied, with notable examples being $\gamma_0 \gamma_5 \gamma_i \nabla_i$, $\gamma_i$, $\epsilon_{ijk}\gamma_j \gamma_k$ and $\nabla_i$ where the fractional overlaps increase by factors of approximately $1.25$, $1.13$, $1.2$ and $1.56$ respectively. The case of $\nabla_i$ is also particularly special since the fractional overlap goes from being below $0.5$ with standard distillation, which one might consider small, to around $0.742$ with the optimal profile, which is around the values that other operators have for the standard distillation case. While this shows that this operator has significant contributions from excited states compared to others of the same $J^{PC}$ and therefore might not be the best one to use to access the ground state, it also holds that the use of the optimal profile significantly decreases such contributions and puts the operator in approximately equal standing to the others with standard distillation. The case of the operators involving $\mathbb{B}_i$ also display an interesting behaviour. Both $\gamma_i \mathbb{B}_i$ and $\gamma_5 \mathbb{B}_i$ present a rather small increase in their fractional overlaps when the optimal profiles are used. This could be explained by the fact that these operators are expected to have considerable contributions from excited states, including members of hybrid supermultiplets with the same $J^{PC}$ \cite{Liu2012}, given the combination of derivatives they contain. Because of this the optimal profile might not be able to suppress these contributions significantly or otherwise enhance too much the one from the ground state already present. However, the contribution to these excited states is expected to be enhanced by the optimal meson profile corresponding to them which is of course different than the one of the ground state.

\subsection{Using quark-connected and quark-disconnected correlations}
\begin{figure}[H]
\centering
\includegraphics[width=0.9\linewidth]{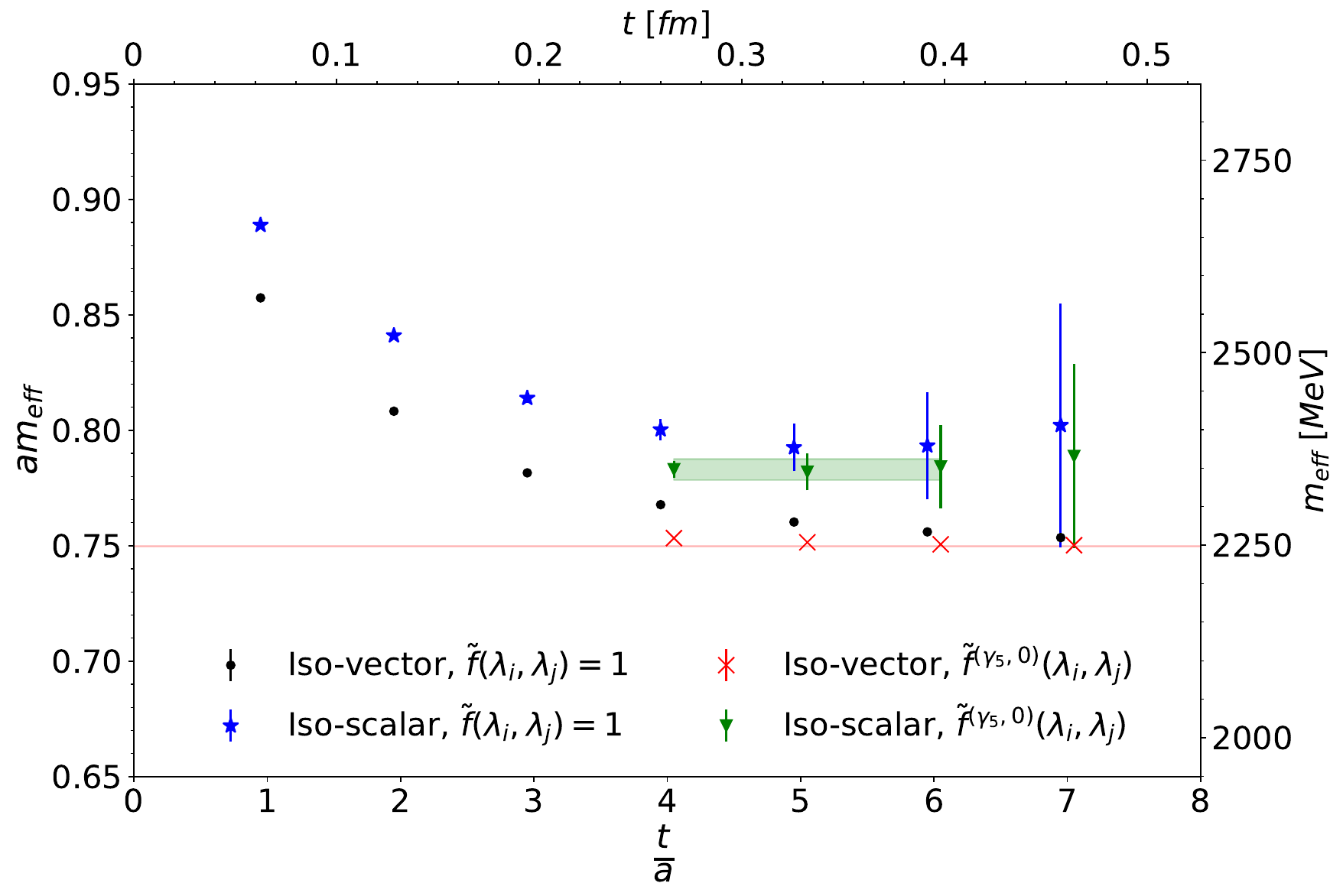}
\caption{Effective mass of the ground state for the iso-scalar and iso-vector $\gamma_5$ operators using standard distillation, $\tilde{f}(\lambda_i, \lambda_j) = 1$, and the optimal profile built from the iso-vector data, $\tilde{f}^{(\gamma_5,0)}(\lambda_i, \lambda_j)$. Data points are displaced horizontally for clarity.}
\label{fig:g5Scalar&Vector}
\end{figure}

\noindent The use of distillation also grants access to the quark-disconnected correlations essential for iso-scalar spectroscopy, seen as the second term in Eq. \eqref{eqn:Meson2Point}. To repeat the GEVP analysis 
to find the optimal distillation profiles requires a calculation of these disconnected pieces.
However the presence of significant noise in these quark-disconnected correlations represents a major problem for the GEVP both in terms of numerical stability and the small number of points that contain a discernible signal. To avoid such complications the optimal profiles for the iso-vector operators are used to build corresponding iso-scalar operators. Two reasons motivate such a choice: first, if the iso-vector and iso-scalar states are not too different in mass then it is not unreasonable to expect their optimal meson distillation profiles might also be not too different. 
Second, even if thet are not very similar, the iso-vector profile might 
still constitute an improvement over the constant profile of standard distillation. With these considerations in mind, all optimal profiles for different operators $\Gamma$ described in this section keep the label $\tilde{f}^{(\Gamma,e)}(\lambda_i,\lambda_j)$ and it is understood that they correspond to those used for the iso-vector analysis. The first operator analyzed is $\Gamma = \gamma_5$. Fig. \ref{fig:g5Scalar&Vector} shows the resulting effective masses for both iso-scalar and iso-vector operators when using both standard distillation and the optimal meson profile of the iso-vector operator. Results corresponding to the optimal profiles are plotted starting at $t_G+a$ and only shown when the error is smaller than the signal. Both profiles display a non-negligible mass difference between iso-scalar and iso-vector channels at early enough times where the error is manageable. 
The use of the optimal profile leads to a considerable reduction of excited state contamination compared to the constant profile. This confirms the intuition that a reasonable profile is a better choice than the constant. From these improved mass data for the iso-scalar one can establish a rather early plateau whose average has a difference of $99\pm 14$ MeV with respect to the improved iso-vector mass plateau average. Initial estimates of this quantity have pointed to relatively small values in lattice studies \cite{ForcrandDisc, McNeileDisc, Levkova2011} and perturbative NRQCD \cite{Hatton2020, Follana, BraatenNRQCD, Adkins, Karplus}, albeit with conflicting signs. Recent lattice calculations of this splitting, either indirectly \cite{Hatton2020} or directly \cite{zhang2021glueball, zhang2021annihilation}, also point to small-yet-positive values. In both cases the value was significantly lower, albeit at a much higher quark mass close to its physical value for the charm quark. Other possible reasons for this disagreement include different flavor content, finite lattice spacing, model assumptions and residual contamination from excited states. Nonetheless it is still clear that the use of an optimal meson profile improves the quality of the result obtained for this mass splitting.

\begin{figure}[H]
\centering
\includegraphics[width=0.9\linewidth]{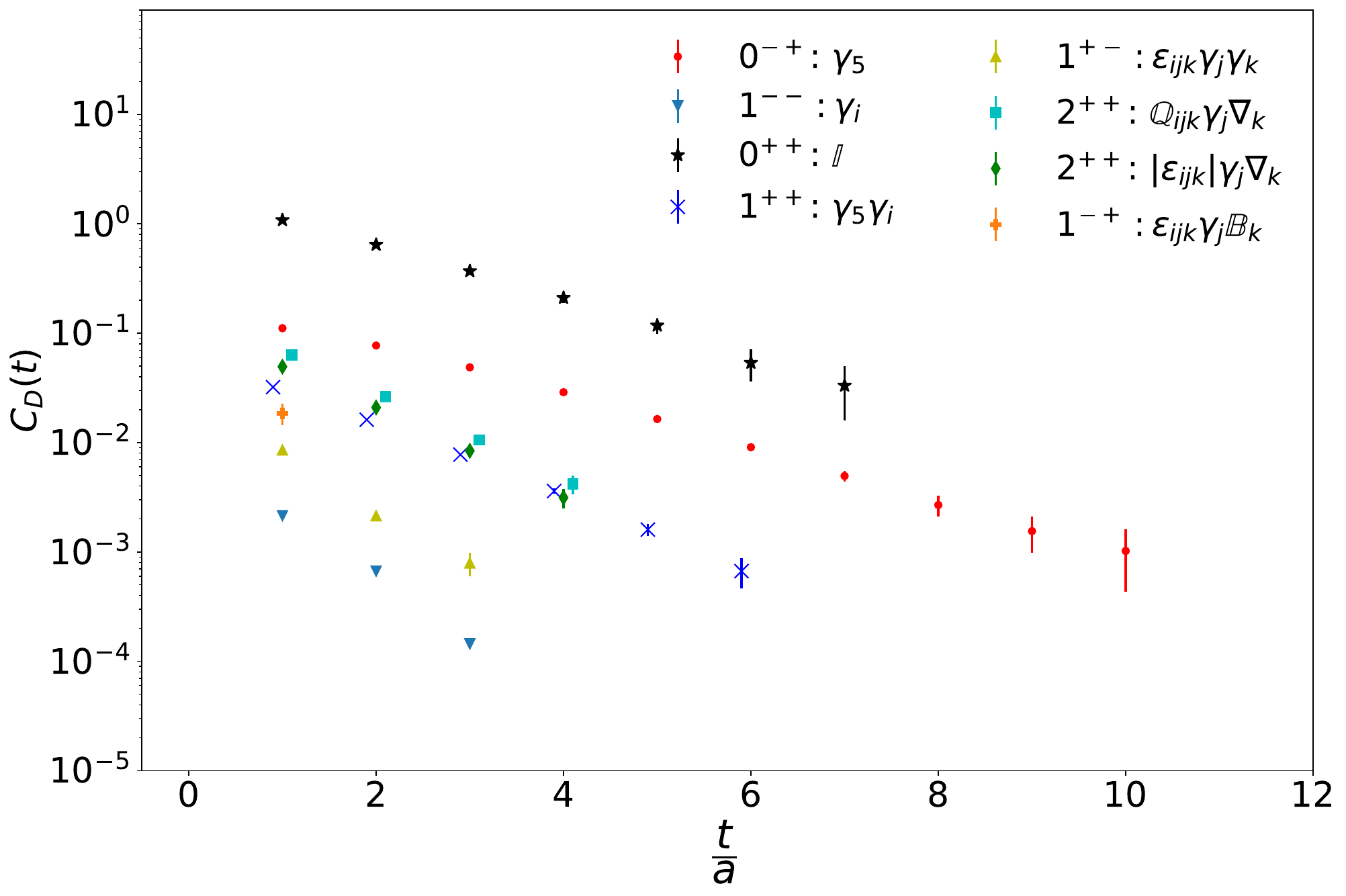}
\caption{Quark-disconnected piece of the correlation of the ground state for some of the operators studied using the profiles $f^{(\Gamma,0)}(\lambda_i, \lambda_j)$. Data points are displaced horizontally for clarity and omitted if the signal is lost to the noise.}
\label{fig:Disconnected_Set1}
\end{figure}

\noindent The disconnected pieces for all other operators used were calculated to assess two main aspects: the presence of a signal at small time-separations and how much statistical noise it has. These dictate the feasability of calculating the correlators of iso-scalar operators. Fig. \ref{fig:Disconnected_Set1} shows the disconnected pieces for the operators that display the best signal when using the corresponding optimal iso-vector meson distillation profiles. Results are shown until the error becomes larger than the signal. The clear signal for the $\gamma_5$ operator is shown in Fig. \ref{fig:g5Scalar&Vector} and serves as evidence for the advantage of using distillation.

\begin{figure}[H]
\centering
\includegraphics[width=0.9\linewidth]{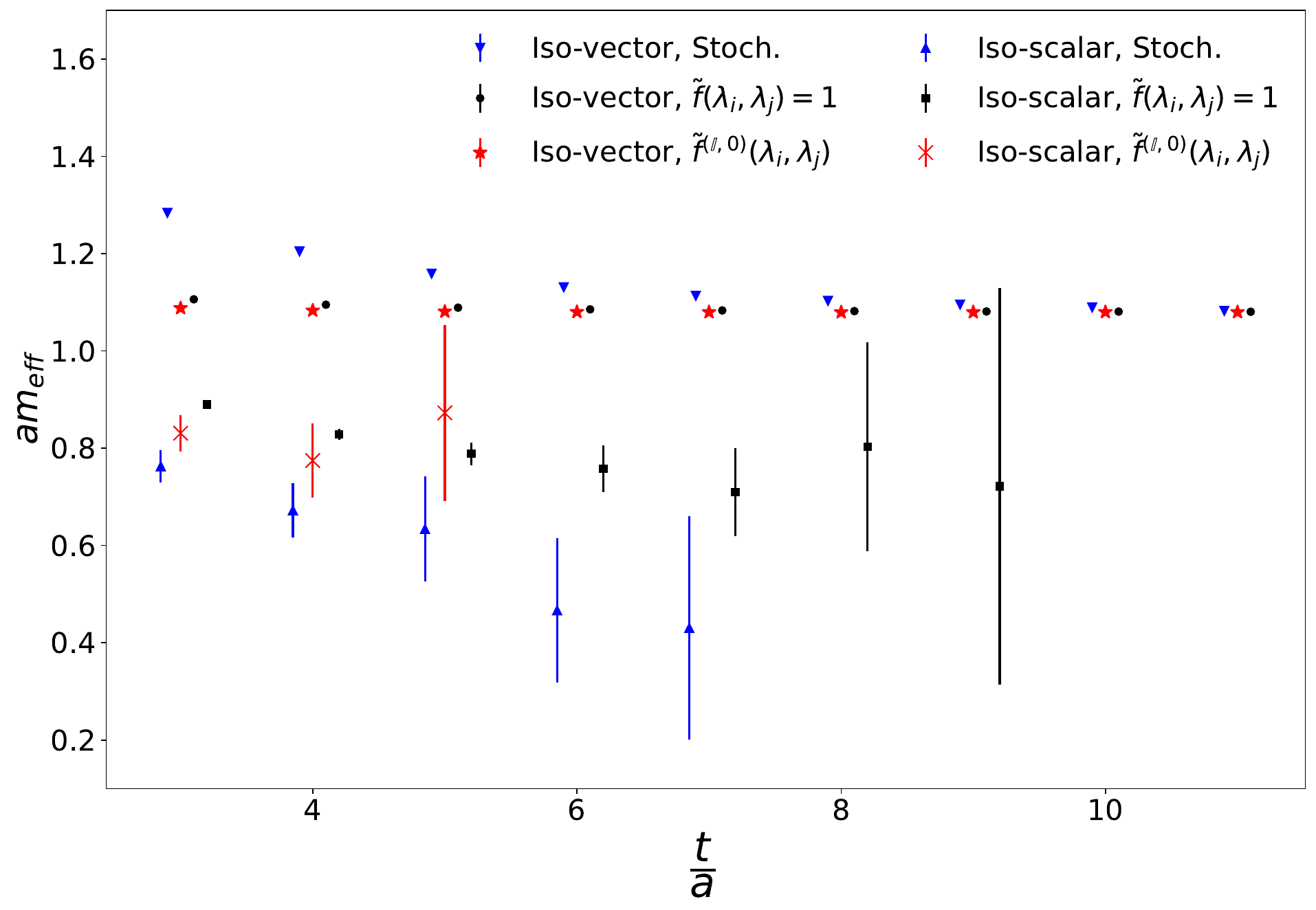}
\caption{Effective masses extracted from the iso-vector and iso-scalar operators defined by $\Gamma= \mathbb{I}$ using standard distillation ($\tilde{f}(\lambda_i,\lambda_j) = 1$), distillation with the optimal profile obtained from the iso-vector data ($\tilde{f}^{(\mathbb{I},0)}(\lambda_i, \lambda_j)$) and stochastic trace estimation. Data points are displaced horizontally for clarity and omitted if the signal is lost to the noise.}
\label{fig:0pp_IsoVector&IsoScalar}
\end{figure}

\noindent 
The clear signal for the disconnected correlation function for operator $\mathbb{I}$ seen in Fig. \ref{fig:Disconnected_Set1} motivates a more detailed study of the iso-scalar $J^{PC}=0^{++}$ channel.  Fig. \ref{fig:0pp_IsoVector&IsoScalar} shows the effective masses for the $0^{++}$ iso-vector and iso-scalar operators using standard distillation, distillation with the optimal profile obtained from the iso-vector data only and operators built from unsmeared quark fields evaluated via stochastic trace estimation. The traces
\begin{align}
\sum_{\vec{x}} \Tr\left[D^{-1}(x,x)\Gamma \right]
\end{align}
required for the latter method are evaluated on every time-slice using 64 $U(1)$ noise vectors per time slice and configuration. $\Phi^{(\Gamma)}$ for all $\Gamma$ matrices can be evaluated using the same inversions. Iso-vector correlators built from unsmeared and from both types of distilled operators lead to the same effective mass at large temporal separations, with distillation using the optimal profile performing considerably better. Meanwhile, the iso-scalar effective masses tend to values significantly lower than the iso-vector case. Not only this, but the use of the optimal profile from the iso-vector for the iso-scalar operator considerably increases the noise. These two observations are not independent. The first indicates that there exists an iso-scalar state, probably a glueball, with quantum numbers $0^{++}$ which is much lighter than the iso-vector $0^{++}$ predominantly created by quark-anti-quark excitation. This explains the second fact: such a mass difference indicates the optimal profile for the heavier state is not sufficiently close to the true optimal profile of the lighter one and therefore is a bad choice to build an operator to create this lighter state. A better approach would include glueball operators together with the iso-scalar $\Gamma = \mathbb{I}$ in the GEVP but this is beyond the scope of the current work.

\begin{figure}[H]
\centering
\includegraphics[width=0.9\linewidth]{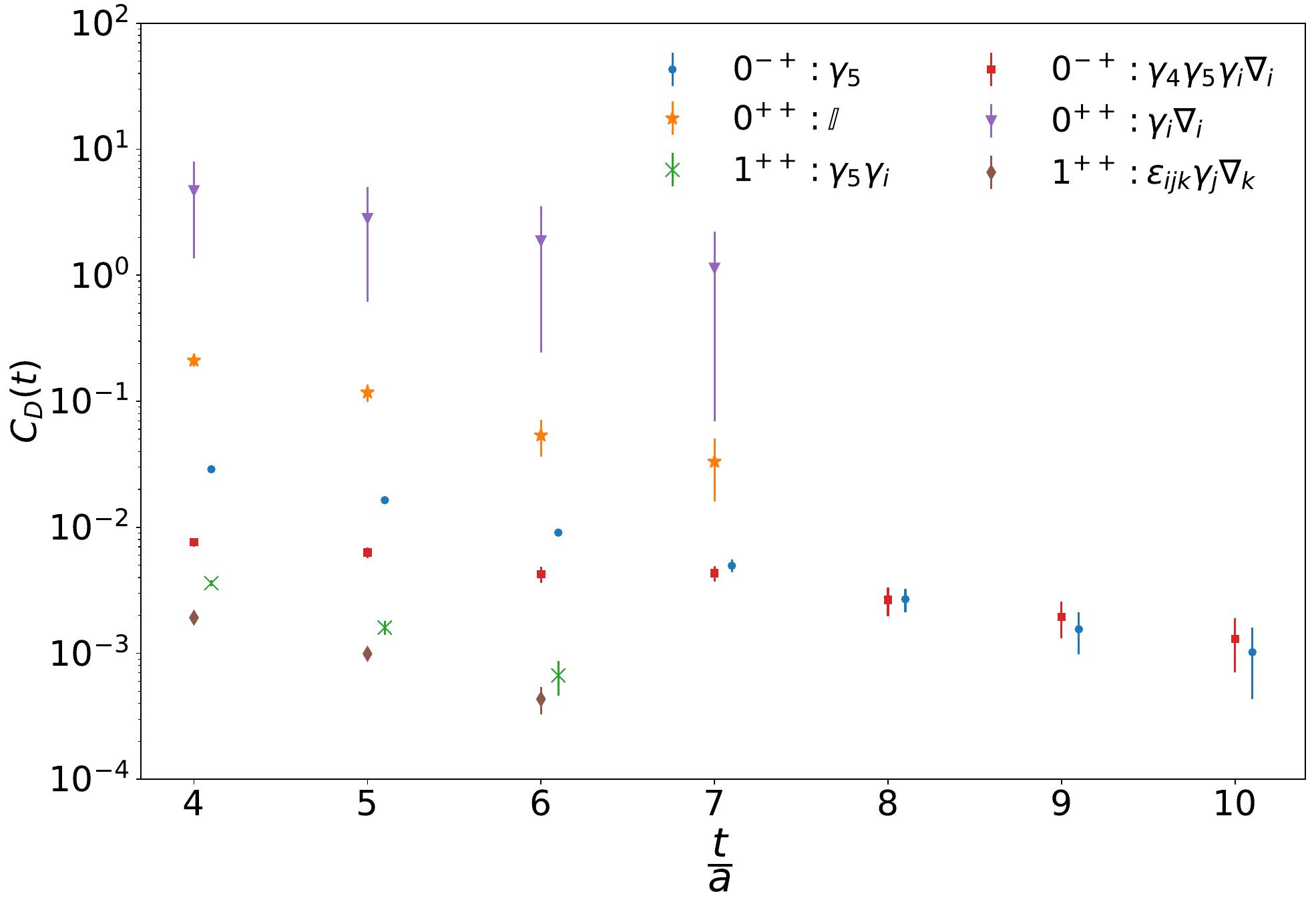}
\caption{The quark-disconnected correlation function of the ground-state for some of the local and derivative-based operators with fixed $J^{PC}$ studied in this work using the profiles $f^{(\Gamma,0)}(\lambda_i, \lambda_j)$. Data points where the signal is lost to the noise are omitted.}
\label{fig:Disc_Local&Derivative}
\end{figure}

\noindent It is also useful to compare the signal of the disconnected correlation obtained with local operators with the derivative-based operators. Since this data has the largest noise, the operators which reduce this noise are the natural choice. a priori, it is not know if the local or derivative-based operators will perform best. Fig. \ref{fig:Disc_Local&Derivative} shows the disconnected correlation for the $0^{++}$, $0^{-+}$ and $1^{++}$ channels using both types with their corresponding profile $\tilde{f}^{(\Gamma,0)}(\lambda_i,\lambda_j)$. 
These channels exhibit the best signal for this correlation. It is clear that both types show a significant signal with only the $0^{++}$ case where there is a major difference in noise between local and derivative operators while the other two channels exhibit somewhat similar errors. This indicates the derivative based operators can also be used to analyze iso-scalar operators reliably. Since the quark-disconnected correlation functions of derivative-based operators were analyzed on a smaller subset of gauge configurations a robust test would first match the statistics of the calculation with local operators. This not only improves the signal for the $\gamma_i \nabla_i$ compared to the $\mathbb{I}$ operator but also for the $0^{-+}$ and $1^{++}$ operators. Since the last two already seem to match the error from local operators, an increase of statistics can only provide better results, which could point to promising results when mixing these different types of operators together.

\section{Conclusions and outlook}
\label{sec:Conclusions}

\noindent In this work, the distillation framework for quark-field smearing in
lattice QCD was extended and techniques for optimising the resulting creation 
operators were developed and tested. The extension applies arbitrary scalar 
functions of each of the eigenvalues of the gauge-covariant laplacian on each 
time-slice inside the smearing kernel. As a consequence of this extension, 
meson profile functions are also introduced and investigated. The procedure 
to build optimal meson distillation profiles via the GEVP formulation was
presented explicitly, making the best use of the Laplace eigenvectors which
form the basis for the distillation method. 
Although only meson operators were examined in this work, the method can be 
extended naturally to other hadrons such as baryons and tetraquarks. Tests were
performed in QCD with $\Nf=2$ degenerate quarks, with a mass close to half 
that of the physical charm-quark. Results for the spectrum of the 
quark-anti-quark system were determined using local and derivative-based 
operators and the comparison between standard distillation and the optimisation
proposed here was made.

\noindent Several advantages were observed. First, comparing the effective mass plots obtained with standard distillation to the optimal profiles, a remarkable decrease in excited state contributions is seen. This effect is graphically seen in Figs. \ref{fig:EffMass_Comparison} and \ref{fig:EffMass_Comparison_Deriv} for some operators and quantified in Table \ref{table:OverlapTable} for all operators considered in this work. Only some of the operators involving a chromomagnetic component exhibit a small improvement, serving as indication that other aspects not influenced by the distillation profile should be improved, such as their gluonic excitations. The increased suppression of excited state contamination in all others yields earlier plateaus with at least five more data points in several cases. Additionally, even where the mass plateaus have equal length those generated from the optimal profiles occur earlier, benefiting studies of quark-disconnected correlation functions. These notoriously suffer from a signal-to-noise problem which sets in at early times. So achieving a mass plateau at early times for the quark-connected correlations gives more precise information to combine with the quark-disconnected correlation. Second, this improvement does not significantly increase the computational work since no additional solutions of the Dirac equation or lattice derivatives must be calculated beyond those required for the perambulator and standard elementals. Third, the appropriate choice of the number of eigenvectors which is crucial to distillation can be approached more systematically. Should too few be chosen, not enough useful information is kept and the resolution is poor. With too many, insufficient quark-field smearing is applied and the computational cost increases. The optimal distillation profiles studied here provide a very useful guide; eigenvectors that are important for a given state have a large contribution in the profile while less important ones are suppressed. 
For a fixed distillation-space size, eigenvectors beyond those included might contain important structure that is subsequently neglected. Introducing profile
functions does not correct for this defect however it does maximally use the 
data in the basis of smooth fields and so can make up in part for the limited 
resolution given by a fixed number of modes. 
The use of too few eigenvectors is detected by monitoring whether the profile is small for the largest eigenvalue included. At the same time the contributions
of higher modes in a very large basis is regulated automatically via the 
optimal distillation profile. Fourth, the optimal meson distillation profiles can be used to visualize the spatial distributions of the resulting meson operators. These distributions provide not only an estimate of the physical extent of the mesons but also suggest if there are notable finite-volume effects when the spatial distribution has sizable contributions near the edges of the lattice. Fifth and final, this improvement is in principle applicable in the stochastic distillation framework, where the perambulators are not calculated exactly. In this case the noise introduced by the estimation may affect the GEVP formulation and therefore the accuracy of the determined profiles. It could be preferable to perform a small statistics, exact calculation of the perambulators to determine the optimal meson profiles for use in a full-statistics study.\\ \\

\noindent A natural next step is to extend the GEVP basis by not only varying the quark distillation profiles of a single fixed $\Gamma$ structure but also including multiple structures, each with a different profile. This should allow not only access to further excited states but also to potentially improve the signal obtained for the ground and first excited states. As pointed out in Sec. \ref{sec:CharmoniumSpectrum} from the results displayed in Fig. \ref{fig:0pp_IsoVector&IsoScalar}, a further step is to build a GEVP basis of iso-scalar operators resembling both mesons and glueballs to explore their possible mixing. Such studies are under way, both using the ensembles studies in this work and with a finer lattice spacing and larger volume to explore the effects of the lattice spacing.\\ \\ 

\textbf{Acknowledgements.} The authors gratefully acknowledge the Gauss Centre for Supercomputing e.V. (\url{www.gauss-centre.eu}) for supporting this project with computing time on the GCS Supercomputer SuperMUC-NG at Leibniz Supercomputing Centre (\url{www.lrz.de}). The work is supported by the LatticeHadrons network
of STRONG-2020 (\url{www.strong-2020.eu}), funded by the European Union Horizon 2020 research and innovation programme under Grant Agreement No. 824093. The authors also acknowledge fruitful discussions with Andreas Frommer, Karsten Kahl, Gustavo Ramirez and Artur Strebel within the DFG research unit FOR5269 "Future methods for studying confined gluons in QCD" and thank Roman Höllwieser for help tuning the APE parameters used in this work.

\begin{appendices}
\numberwithin{equation}{section}
\clearpage

\section{Optimal $\tilde{\Gamma}$ structure}
\label{app:OptimalGamma}

To define an optimal $\tilde{\Gamma}^{(\Gamma,e)}$ structure that includes the optimal meson distillation profile Eq. \eqref{eqn:OptimalElementalMatrix} at meson level one starts with the original $\Gamma$ structure in the channel of interest. Assuming the decomposition into an operator $\mathcal{H}$ acting on Dirac space and an operator $\mathcal{D}[t]$ acting on coordinate/color space, one can perform a change of basis to express $\mathcal{D}[t]$ in terms of $V_{N}[t]$, a matrix whose columns are all the Laplacian eigenvectors at time $t$. Doing this results in
\begin{align}
\Gamma[t] &= \mathcal{H} \mathcal{D}[t]\\
&= \mathcal{H} V_N[t] V_N[t]^{\dagger} \mathcal{D}[t] V_N[t] V_N[t]^{\dagger}\\
&= \mathcal{H} V_N[t] \Phi_N[t] V_N[t]^{\dagger},
\end{align}
where $\Phi_N[t]$ corresponds to the elemental of $\mathcal{D}[t]$ when all eigenvectors are known. If one now considers the meson operator
\begin{equation}
\mathcal{O}(t) = \bar{q}(t) \Gamma q(t)
\end{equation}
it corresponds to applying standard distillation to the quark fields using all of the Laplacian eigenvectors, which is known to have no effect. However, one can exploit this formulation of the $\Gamma$ structure to manually insert at meson level an arbitrary meson distillation profile just as one could introduce at quark level the quark distillation profile via the matrix $J[t]$. To do this, one replaces $\Phi_N[t]$ with 
\begin{equation}
\tilde{\Phi}_N[t] = \Phi_N[t] \circledast F_N[t]
\end{equation}
where the entries $\left( F_N[t] \right)_{ij} = f(\lambda_i[t], \lambda_j[t])$ include the meson profile $f$ one chooses and $\circledast$ denotes the elementwise product. $\tilde{\Gamma}[t]$ is now given by
\begin{align}
\tilde{\Gamma}[t] = \mathcal{H} V_N[t] \tilde{\Phi}_N[t] V_N[t]^{\dagger}
\end{align}
and a meson operator built using $\tilde{\Gamma}$ is given by
\begin{equation}
\tilde{\mathcal{O}}(t) = \bar{q}(t) \tilde{\Gamma} q(t).
\label{eqn:OriginalMesonOperator}
\end{equation}
The two-point correlation function of such an operator, considering for now only the quark-connected piece for simplicity, will be given by
\begin{align}
C(t) = \left \langle (\pm) \Tr\left( \tilde{\Phi}_N[t] \tau_N[t,0] \tilde{\Phi}_N[0] \tau_N[0,t] \right) \right \rangle,
\end{align}
where the sign depends on $\Gamma$ and $\tau_N[t_1,t_2]$ corresponds to the perambulator involving all eigenvectors. Such a correlation function is not tractable since one would need to know all the eigenvectors and eigenvalues of the Laplacian at every value of time and perform an unfeasable amount of inversions of the Dirac operator. However, one can avoid such a problem by using standard distillation, where the quark fields are replaced by their distilled counterparts using a manageable number of eigenvectors. The distilled meson operator will be given by
\begin{align}
\tilde{\mathcal{O}}_D(t) = \bar{q}(t) V[t] V[t]^{\dagger} \tilde{\Gamma} V[t] V[t]^{\dagger} q(t)
\label{eqn:DistilledMesonOperator}
\end{align}
and the corresponding correlation function is given by
\begin{equation}
C_D(t) = \left \langle (\pm) \Tr\left( \tilde{\Phi}[t] \tau[t,0] \tilde{\Phi}[0] \tau[0,t] \right) \right \rangle,
\end{equation}
where $\tilde{\Phi}[t]$ is the sub-block of $\tilde{\Phi}_N[t]$ corresponding to the chosen lowest eigenpairs and the same for $\tau[t_1,t_2]$ with respect to $\tau_N[t_1,t_2]$. Two observations can be made at this point. First, if one chooses the function that defines $F_N[t]$ to be the optimal meson profiles calculated from the GEVP described in Sec. \ref{sec:SmearProfOpti} then $C_D(t)$ corresponds to the same correlation function that one would obtain  from the eigenvalues of said GEVP. This can be seen first by replacing $C(t)$ with $C_S(t)$ (See Eq. \eqref{eqn:PrunedCorrelation}) in Eq. \eqref{eqn:GEVP} and then multiplying from the left by $w_k(t,t_G)^{\dagger}$, yielding
\begin{align}
\begin{split}
w_k(t,t_G)^{\dagger} C_S(t) w_e(t,t_G) &= \rho_{e}(t,t_G) w_k(t,t_G)^{\dagger}  C_S(t_G) w_e(t,t_G)\\
&= \rho_{e}(t,t_G) \delta_{ke},
\end{split}
\label{eqn:GEVP_Eigenvalue}
\end{align}
where the orthonormality condition 
\beq
w_k(t,t_G)^{\dagger}  C_S(t_G) w_e(t,t_G) = \delta_{e,k}
\eeq
has been used. On the right hand side of Eqn. \eqref{eqn:GEVP_Eigenvalue} are the generalized eigenvalues of the GEVP while on the left hand side is the projected correlation matrix $w_k(t,t_G)^{\dagger} C_S(t) w_e(t,t_G)$. To relate this projected correlation to $C_{D}(t)$ it is important to note that eventhough the vectors $w_e(t,t_G)$ have time indices they are expected to be independent of time up to noise effects. The results presented in this work are built using $w_e(t,t_G)$ for a fixed value of $t$ where there is no significant change between consecutive times so this time independence is assumed and the $t,t_G$ pair is omitted in the following treatment. Expressing the entries of $C(t)$ for a fixed $\Gamma$ as
\beq
C(t)_{mn} = \left \langle \mathcal{O}_m(t) \bar{\mathcal{O}}_n(0) \right \rangle_{F,U}
\eeq
one can express the entries of $C_S(t)$ as
\begin{align}
\begin{split}
C_S(t)_{a,b} &= \left \langle \mathcal{P}_{a}(t) \bar{\mathcal{P}}_{b}(0) \right \rangle_{F,U}
\end{split},
\end{align}
where the pruned operator $\mathcal{P}_{a}(t)$ is given by
\beq
\mathcal{P}_{a}(t) = \sum_{m} u_{a,m}^{*} \mathcal{O}_m(t)
\eeq
and $u_{a,m}$ denotes the $m$-th entry of the $a$-th singular vector used for the pruning. This pruned meson operator can be examined in more detail by explicitly writing the orginal meson operators $\mathcal{O}_m$ in terms of the quak fields. This yields
\begin{align}
\begin{split}
\mathcal{P}_{a}(t) &=\bar{q}(t) V[t] \Phi^{(p,a)}[t] V[t]^{\dagger} q(t)
\end{split},
\end{align}
where the pruned elemental $\Phi^{(p,\Gamma, a)}[t]$ is defined as
\beq
\Phi^{(p,\Gamma ,a)}[t] = \sum_{m} u_{a,m}^{*} J_m[t]^{\dagger} V[t]^{\dagger} \Gamma V[t] J_n[t]
\eeq
and has entries given by
\beq
\Phi^{(p,\Gamma ,a)}_{\substack{\alpha \beta\\ i j}}[t] = \Lambda^{(\Gamma)}_{\substack{\alpha \beta\\ij}}[t] f_a^{(p,\Gamma)}(\lambda_i[t],\lambda_j[t]),
\eeq
where $\Lambda^{(\Gamma)}[t]$ is as defined in Eq. \eqref{eqn:OriginalElemental} and the pruned profile $f_a^{(p,\Gamma)}(\lambda_i[t],\lambda_j[t])$ is as defined in Eq. \eqref{eqn:PrunedProfile} ($u_{a,m}^{*}$ is real). Now the projected correlation for $k=e$ for a fixed $e$ in Eq. \eqref{eqn:GEVP_Eigenvalue} can be written as
\begin{align}
\begin{split}
w_e^{(\Gamma,e)\dagger} C_S(t) w_e^{(\Gamma,e)} &= \left \langle \mathcal{A}^{(\Gamma,e)}(t) \bar{\mathcal{A}}^{(\Gamma,e)}(0) \right \rangle_{F,U}
\end{split}
\label{eqn:ProjectedCorr}
\end{align}
where the operator $\mathcal{A}^{(\Gamma,e)}(t)$ is given by 
\beq
\mathcal{A}^{(\Gamma,e)}(t) = \sum_{a} w^{(\Gamma,e)}_a \mathcal{P}_a(t).
\eeq
It is clear from Eq. \eqref{eqn:ProjectedCorr} that the projected correlation $w_e^{(\Gamma,e)\dagger} C_{S}(t) w_e^{(\Gamma,e)}$ is the correlation of the meson operator $\mathcal{A}^{(\Gamma,e)}$, which is built from the basis of pruned operators $\mathcal{P}_a$ via a linear combination whose coefficients are the entries of the GEVP eigenvector $w_e^{(\Gamma, e)}$. One can now simplify $\mathcal{A}^{(\Gamma,e)}(t)$ as
\begin{align}
\begin{split}
\mathcal{A}^{(\Gamma,e)}(t) &= \bar{q}(t) V[t] \Phi_{\mathcal{A}}^{(\Gamma,e)}[t] V[t]^{\dagger} q(t)
\end{split}
\end{align} 
where the elemental $\Phi_{\mathcal{A}}^{(\Gamma,e)}[t]$ is defined as
\beq
\Phi_{\mathcal{A}}^{(\Gamma,e)}[t] = \sum_{a} w^{(\Gamma,e)}_a \Phi^{(p,\Gamma ,a)}[t]
\eeq
and has entries given as
\beq
\Phi_{\mathcal{A}}^{(\Gamma,e)}[t]_{\substack{ij\\\alpha \beta}} = \Lambda^{(\Gamma)}_{\substack{\alpha \beta\\ij}}[t] \tilde{f}^{(\Gamma,e)}(\lambda_i[t],\lambda_j[t]),
\eeq
where $\tilde{f}^{(\Gamma,e)}(\lambda_i[t],\lambda_j[t])$ is as defined in Eq. \eqref{eqn:OptimalMesonProfile}. Clearly $\Phi_{\mathcal{A}}^{(\Gamma,e)}[t]$ is equal to the optimal meson elemental obtained from the GEVP as given in Eq. \eqref{eqn:OptimalElementalMatrix}, which means that the projected correlation $w_e^{(\Gamma,e)\dagger} C_S(t) w_e^{(\Gamma,e)}$ from the GEVP is equal to $C_D(t)$ since they are built using identical elementals. However, while in the GEVP case the distillation profiles were introduced at quark level independently of the $\Gamma$ structure and different optimal meson distillation profile was obtained for every choice of $\Gamma$ and energy state, here the optimal meson profile is inserted via a redefinition of the $\Gamma$ structure while standard distillation, i.e a constant profile, is used for the quarks.
The second observation is the fact that even though $\tilde{\Gamma}$ cannot be explicitely built without an excessive amount of computational work one can construct an approximation with a limited number of eigenvectors which is given by 
\begin{align}
\tilde{\Gamma}_D[t] &= V[t] V[t]^{\dagger} \tilde{\Gamma} V[t] V[t]^{\dagger}\\
&= \mathcal{H} V[t] \tilde{\Phi}[t] V[t]^{\dagger},
\end{align}
where it can be seen that when all eigenvectors are used one recovers $\tilde{\Gamma}[t]$. The distilled meson operator in Eq. \eqref{eqn:DistilledMesonOperator} can now be written as
\begin{align}
\tilde{O}_D(t) = \bar{q}(t) \tilde{\Gamma}_D q(t)
\end{align}
and corresponds to the closest one can get to the meson operator defined in Eq. \eqref{eqn:OriginalMesonOperator}. Furthermore, if $\tilde{\Gamma}_D$ leads to improved correlation functions, as is the case in this work, then it probably conserves useful properties of $\tilde{\Gamma}$ that can be studied without the need for excessive computational work.

\end{appendices}

\addcontentsline{toc}{section}{References}
\bibliographystyle{JHEP}
\bibliography{bib_improved_distillation}

\providecommand{\href}[2]{#2}\begingroup\raggedright\begin{thebibliography}{10}

\bibitem{Peardon2009:Distillation}
M.~Peardon, J.~Bulava, J.~Foley, C.~Morningstar, J.~Dudek, R.~G. Edwards,
  B.~Jo{\'{o}}, H.-W. Lin, D.~G. Richards, and K.~J. Juge, {\it Novel
  quark-field creation operator construction for hadronic physics in lattice
  {QCD}},  {\em Physical Review D} {\bf 80} (Sept., 2009).

\bibitem{Allton1993:Jacobi}
C.~R. Allton, C.~T. Sachrajda, R.~M. Baxter, S.~P. Booth, K.~C. Bowler,
  S.~Collins, D.~S. Henty, R.~D. Kenway, B.~J. Pendleton, D.~G. Richards, J.~N.
  Simone, A.~D. Simpson, B.~E. Wilkes, and C.~Michael, {\it Gauge-invariant
  smearing and matrix correlators using wilson fermions at$\beta$=6.2},  {\em
  Physical Review D} {\bf 47} (June, 1993) 5128--5137.

\bibitem{Morningstar2011}
C.~Morningstar, J.~Bulava, J.~Foley, K.~J. Juge, D.~Lenkner, M.~Peardon, and
  C.~H. Wong, {\it Improved stochastic estimation of quark propagation with
  laplacian heaviside smearing in lattice {QCD}},  {\em Physical Review D} {\bf
  83} (June, 2011).

\bibitem{Liu2012}
L.~Liu, G.~Moir, M.~Peardon, S.~M. Ryan, C.~E. Thomas, P.~Vilaseca, J.~J.
  Dudek, R.~G. Edwards, B.~Jo{\'{o}}, and D.~G. Richards, {\it Excited and
  exotic charmonium spectroscopy from lattice {QCD}},  {\em Journal of High
  Energy Physics} {\bf 2012} (July, 2012).

\bibitem{Cheung2016}
G.~K. Cheung, , C.~O'Hara, G.~Moir, M.~Peardon, S.~M. Ryan, C.~E. Thomas, and
  D.~Tims, {\it Excited and exotic charmonium, $d_s$ and $d$ meson spectra for
  two light quark masses from lattice {QCD}},  {\em Journal of High Energy
  Physics} {\bf 2016} (Dec., 2016).

\bibitem{Dudek2010}
J.~J. Dudek, R.~G. Edwards, M.~J. Peardon, D.~G. Richards, and C.~E. Thomas,
  {\it Toward the excited meson spectrum of dynamical {QCD}},  {\em Physical
  Review D} {\bf 82} (Aug., 2010).

\bibitem{Dudek2013}
J.~J. Dudek, R.~G. Edwards, P.~Guo, and C.~E. Thomas, {\it Toward the excited
  isoscalar meson spectrum from lattice {QCD}},  {\em Physical Review D} {\bf
  88} (Nov., 2013).

\bibitem{zhang2021glueball}
R.~Zhang, W.~Sun, Y.~Chen, M.~Gong, L.-C. Gui, and Z.~Liu, {\it The glueball
  content of $\eta_c$},  {\em Physics Letters B} {\bf 827} (2022) 136960.

\bibitem{zhang2021annihilation}
R.~Zhang, W.~Sun, F.~Chen, Y.~Chen, M.~Gong, X.~Jiang, and Z.~Liu, {\it
  Annihilation diagram contribution to charmonium masses},  {\em Chinese
  Physics C} {\bf 46} (apr, 2022) 043102.

\bibitem{Blossier2009:GEVP}
{\bf ALPHA} Collaboration, B.~Blossier, M.~D. Morte, G.~von Hippel, T.~Mendes,
  and R.~Sommer, {\it On the generalized eigenvalue method for energies and
  matrix elements in lattice field theory},  {\em Journal of High Energy
  Physics} {\bf 2009} (Apr., 2009) 094--094.

\bibitem{Luscher1990}
M.~L\"{u}scher and U.~Wolff, {\it How to calculate the elastic scattering
  matrix in two-dimensional quantum field theories by numerical simulation},
  {\em Nuclear Physics B} {\bf 339} (July, 1990) 222--252.

\bibitem{Irges2007}
N.~Irges and F.~Knechtli, {\it Lattice gauge theory approach to spontaneous
  symmetry breaking from an extra dimension},  {\em Nuclear Physics B} {\bf
  775} (July, 2007) 283--311.

\bibitem{Balog1999}
J.~Balog, M.~Niedermaier, F.~Niedermayer, A.~Patrascioiu, E.~Seiler, and
  P.~Weisz, {\it Comparison of the o(3) bootstrap $\sigma$ model with lattice
  regularization at low energies},  {\em Physical Review D} {\bf 60} (Oct.,
  1999).

\bibitem{Niedermayer2001}
F.~Niedermayer, P.~R\"{u}fenacht, and U.~Wenger, {\it Fixed point gauge actions
  with fat links: scaling and glueballs},  {\em Nuclear Physics B} {\bf 597}
  (Mar., 2001) 413--450.

\bibitem{Jansen:1998mx}
{\bf ALPHA} Collaboration, K.~Jansen and R.~Sommer, {\it {O(a) improvement of
  lattice QCD with two flavors of Wilson quarks}},  {\em Nucl. Phys. B} {\bf
  530} (1998) 185--203. [Erratum: Nucl.Phys.B 643, 517--518 (2002)].

\bibitem{Fritzsch:2012wq}
P.~Fritzsch, F.~Knechtli, B.~Leder, M.~Marinkovic, S.~Schaefer, R.~Sommer, and
  F.~Virotta, {\it {The strange quark mass and Lambda parameter of two flavor
  QCD}},  {\em Nucl. Phys. B} {\bf 865} (2012) 397--429.

\bibitem{Cali:2019enm}
S.~Cal{\`i}, F.~Knechtli, and T.~Korzec, {\it {How much do charm sea quarks
  affect the charmonium spectrum?}},  {\em Eur. Phys. J. C} {\bf 79} (2019),
  no.~7 607.

\bibitem{Luscher:2010iy}
M.~L\"uscher, {\it {Properties and uses of the Wilson flow in lattice QCD}},
  {\em JHEP} {\bf 08} (2010) 071. [Erratum: JHEP 03, 092 (2014)].

\bibitem{SorensenChebyshev}
D.~C. Sorensen and C.~Yang, {\it Accelerating the lanczos algorithm via
  polynomial spectral transformations},  tech. rep., Rice University, 1997.

\bibitem{Wu2000}
K.~Wu and H.~Simon, {\it Thick-restart lanczos method for large symmetric
  eigenvalue problems},  {\em {SIAM} Journal on Matrix Analysis and
  Applications} {\bf 22} (Jan., 2000) 602--616.

\bibitem{GrcarPeriodic}
J.~F. Grcar, {\em Analyses of the Lanczos Algorithm and of the Approximation
  Problem in Richardson's Method}.
\newblock PhD thesis, University of Illinois at Urbana-Champaign, 1981.

\bibitem{Simon1984}
H.~D. Simon, {\it Analysis of the symmetric lanczos algorithm with
  reorthogonalization methods},  {\em Linear Algebra and its Applications} {\bf
  61} (Sept., 1984) 101--131.

\bibitem{Greensite2005}
J.~Greensite, {\v{S}}.~Olejn{\'{\i}}k, M.~I. Polikarpov, S.~N. Syritsyn, and
  V.~I. Zakharov, {\it Localized eigenmodes of covariant laplacians in the
  yang-mills vacuum},  {\em Physical Review D} {\bf 71} (June, 2005).

\bibitem{Albanese1987}
M.~Albanese, F.~Costantini, G.~Fiorentini, F.~Flore, M.~Lombardo,
  R.~Tripiccione, P.~Bacilieri, L.~Fonti, P.~Giacomelli, E.~Remiddi,
  M.~Bernaschi, N.~Cabibbo, E.~Marinari, G.~Parisi, G.~Salina, S.~Cabasino,
  F.~Marzano, P.~Paolucci, S.~Petrarca, F.~Rapuano, P.~Marchesini, and
  R.~Rusack, {\it Glueball masses and string tension in lattice {QCD}},  {\em
  Physics Letters B} {\bf 192} (June, 1987) 163--169.

\bibitem{Luscher2013}
M.~L\"{u}scher and S.~Schaefer, {\it Lattice {QCD} with open boundary
  conditions and twisted-mass reweighting},  {\em Computer Physics
  Communications} {\bf 184} (Mar., 2013) 519--528.

\bibitem{Frommer:2013fsa}
A.~Frommer, K.~Kahl, S.~Krieg, B.~Leder, and M.~Rottmann, {\it {Adaptive
  Aggregation Based Domain Decomposition Multigrid for the Lattice Wilson Dirac
  Operator}},  {\em SIAM J. Sci. Comput.} {\bf 36} (2014) A1581--A1608.

\bibitem{Wolff2004}
U.~Wolff, {\it Monte carlo errors with less errors},  {\em Computer Physics
  Communications} {\bf 156} (Jan., 2004) 143--153.

\bibitem{Schaefer2011}
S.~Schaefer, R.~Sommer, and F.~Virotta, {\it Critical slowing down and error
  analysis in lattice {QCD} simulations},  {\em Nuclear Physics B} {\bf 845}
  (Apr., 2011) 93--119.

\bibitem{Dudek2008Operators}
J.~J. Dudek, R.~G. Edwards, N.~Mathur, and D.~G. Richards, {\it Charmonium
  excited state spectrum in lattice {QCD}},  {\em Physical Review D} {\bf 77}
  (Feb., 2008).

\bibitem{Burch2009}
T.~Burch, C.~Hagen, M.~Hetzenegger, and A.~Sch\"{a}fer, {\it Low and high spin
  mesons {fromNf}=2clover-wilson lattices},  {\em Physical Review D} {\bf 79}
  (June, 2009).

\bibitem{Gattringer2008}
C.~Gattringer, L.~Y. Glozman, C.~B. Lang, D.~Mohler, and S.~Prelovsek, {\it
  Derivative sources in lattice spectroscopy of excited light-quark mesons},
  {\em Physical Review D} {\bf 78} (Aug., 2008).

\bibitem{Dudek2009}
J.~J. Dudek, R.~G. Edwards, M.~J. Peardon, D.~G. Richards, and C.~E. Thomas,
  {\it Highly excited and exotic meson spectrum from dynamical lattice {QCD}},
  {\em Physical Review Letters} {\bf 103} (Dec., 2009).

\bibitem{UrreaProc}
J.~A. Urrea-Niño, F.~Knechtli, T.~Korzec, and M.~Peardon, {\it Optimizing
  distillation for charmonium and glueballs},  in {\em Proceedings of The 38th
  International Symposium on Lattice Field Theory {\textemdash}
  {PoS}({LATTICE}2021)}, Sissa Medialab, may, 2022.

\bibitem{MankeOperators}
X.~Liao and T.~Manke, {\it Excited charmonium spectrum from anisotropic
  lattices},  2002.

\bibitem{DeTar2019}
C.~DeTar, A.~S. Kronfeld, S.~haeng Lee, D.~Mohler, and J.~N.~S. and, {\it
  Splittings of low-lying charmonium masses at the physical point},  {\em
  Physical Review D} {\bf 99} (Feb., 2019).

\bibitem{ForcrandDisc}
T.~Q.-T. C.~P. Forcrand, M.~G. a~P~rez, H.~Matsufuru, A.~Nakamura, I.~Pushkina,
  I.-O. Stamatescu, T.~Takaishi, and T.~Umeda, {\it Contribution of
  disconnected diagrams to the hyperfine splitting of charmonium},  {\em
  Journal of High Energy Physics} {\bf 2004} (aug, 2004) 004--004.

\bibitem{McNeileDisc}
{\bf UKQCD Collaboration} Collaboration, C.~McNeile and C.~Michael, {\it
  Estimate of the flavor singlet contributions to the hyperfine splitting in
  charmonium},  {\em Phys. Rev. D} {\bf 70} (Aug, 2004) 034506.

\bibitem{Levkova2011}
L.~Levkova and C.~DeTar, {\it Charm annihilation effects on the hyperfine
  splitting in charmonium},  {\em Physical Review D} {\bf 83} (Apr., 2011).

\bibitem{Hatton2020}
D.~Hatton, C.~Davies, B.~Galloway, J.~Koponen, G.~Lepage, and A.~L. and, {\it
  Charmonium properties from lattice {QCD}+{QED} : Hyperfine splitting,
  $j/\psi$ leptonic width, charm quark mass, and $a_{\mu}^c$},  {\em Physical
  Review D} {\bf 102} (Sept., 2020).

\bibitem{Follana}
E.~Follana, Q.~Mason, C.~Davies, K.~Hornbostel, G.~P. Lepage, J.~Shigemitsu,
  H.~Trottier, and K.~Wong, {\it Highly improved staggered quarks on the
  lattice with applications to charm physics},  {\em Phys. Rev. D} {\bf 75}
  (Mar, 2007) 054502.

\bibitem{BraatenNRQCD}
G.~T. Bodwin, E.~Braaten, and G.~P. Lepage, {\it Rigorous qcd analysis of
  inclusive annihilation and production of heavy quarkonium},  {\em Phys. Rev.
  D} {\bf 51} (Feb, 1995) 1125--1171.

\bibitem{Adkins}
G.~S. Adkins, {\it Application of the bound state formalism to positronium},
  {\em AIP Conference Proceedings} {\bf 189} (1989), no.~1 65--92.

\bibitem{Karplus}
R.~Karplus and A.~Klein, {\it Electrodynamic displacement of atomic energy
  levels. iii. the hyperfine structure of positronium},  {\em Phys. Rev.} {\bf
  87} (Sep, 1952) 848--858.

\end{thebibliography}\endgroup

\end{document}